\documentclass[aps,prc,reprint,groupedaddress]{revtex4-1}
\usepackage{amsmath}
\usepackage{graphicx}
\usepackage{array}
\usepackage{capt-of}
\usepackage{float}
\usepackage{placeins}

\begin{document}
\newcommand{\PIEL}{$\pi N \rightarrow \pi N $ }	
\newcommand{\GPPN}{$\gamma N \rightarrow \pi N $ }
\newcommand{\GPEP}{$\gamma p \rightarrow \eta p $ }
\newcommand{\GPKL}{$\gamma p \rightarrow K^+ \Lambda $ }
\newcommand{\GNKL}{$\gamma n \rightarrow K^0 \Lambda $ }
\newcommand{\GNEN}{$\gamma n \rightarrow \eta n $ }
\newcommand{\PINE}{$\pi^- p \rightarrow \eta n$}
\newcommand{\PINK}{$\pi^- p \rightarrow K^0 \Lambda$}
\newcommand{\DSG}{$d\sigma / d \Omega$}

\newcommand{\trimAA}{3mm}
\newcommand{\trimBB}{10mm}
\newcommand{\trimCC}{15mm}
\newcommand{\trimDD}{19mm}

\newcommand{\pathGPEP}{Figures/GPEP/}
\newcommand{\pathGNEN}{Figures/GNEN/}
\newcommand{\pathGPKL}{Figures/GPKL/}
\newcommand{\pathGNKL}{Figures/GNKL/}
\newcommand{\pathGPPIP}{Figures/GPPIP/}
\newcommand{\pathGNPIP}{Figures/GNPIP/}
\newcommand{\pathPINE}{Figures/PINE/}
\newcommand{\pathPINK}{Figures/PINK/}
\newcommand{\DSGT}{$\frac{d\sigma}{d\Omega}$}

\newcommand{\HSP}{H_{SP}}
\newcommand{\HSA}{H_{SA}}
\newcommand{\HN}{H_{N}}
\newcommand{\HD}{H_{D}}
\newcommand{\HSPC}{H_{SP}^*}
\newcommand{\HSAC}{H_{SA}^*}
\newcommand{\HNC}{H_{N}^*}
\newcommand{\HDC}{H_{D}^*}
\newcommand{\qk}{\frac{q}{k}}
\newcolumntype{C}{>{ \arraybackslash}p{7em}}

\newcommand{\bibA}[4]{#1, #2 \textbf{#3}, #4}
\newcommand{\bibE}[4]{#1 \textit{et al.}, #2 \textbf{#3}, #4.}
\newcommand{\PRL}{Phys. Rev. Lett.}

\newcommand{\printObs}{\DSGT}
\newcommand{\printReac}{\GPEP}
\newcommand{\ObsName}{DSG}
\newcommand{\pathGraphing}{\pathGPEP}



\title{Partial-Wave Analysis of $\gamma p \rightarrow K^+ \Lambda$ using a multichannel framework}


	\author{B. C. Hunt and D. M. Manley}
	\affiliation{Department of Physics, Kent State University, Kent, OH 44242-0001, USA}
	

\date{\today}

\begin{abstract}
 Results from a partial-wave analysis of the reaction \GPKL \ are presented. The reaction is dominated by the $S_{11}(1650)$ and $P_{13}(1720)$ resonances at low energies and by $P_{13}(1900)$ at higher energies. There are small contributions from all amplitudes up to and including $G_{17}$, with $F_{17}$ necessary for obtaining a good fit of several of the spin observables. We find evidence for $P_{11}$(1880), $D_{13}$(2120), and $D_{15}$(2080) resonances, as well as a possible $F_{17}$ resonance near 2300~MeV, which is expected from quark-model predictions.  Some predictions for $\gamma n \to K^0 \Lambda$ are also included.
\end{abstract}

\pacs{}

\maketitle

\section{Introduction}
Quark models predict a larger number of resonances than have been experimentally verified. This is called the problem of the ``missing resonances.'' One possible explanation is that these resonances decouple from the $\pi N$ channel. Photoproduction reactions involving final states other than $\pi N$ provide an opportunity to search for these resonances. 

One of the best measured of these reactions is $\gamma p \to K^+ \Lambda$. Of the 16 single- and double-polarization observables, eight have been measured and published while the others are in some stage of analysis. With measurements of so many observables, it might be expected that only a single unique solution is permitted by the data; however, there are still major discrepancies between the amplitudes determined by different groups. One encouragement is that the predicted resonance spectrum is in good agreement. This paper is laid out as follows: Sec.~II describes the formalism and methodology for the partial-wave analysis, Sec.~III describes the procedure used, Sec.~IV describes results from the analysis, and Sec.~V presents a summary and conclusions.  Finally, Appendix~A compares our final energy-dependent solution with the data included in our analysis.

\section{Formalism and Methodology}
Four helicity amplitudes describe the photoproduction of a pseudoscalar ($J^P=0^-$) meson and a $J^P=\frac{1}{2}^+$ baryon off of a nucleon target \cite{Walker}. Each of the four helicity amplitudes can be expanded in terms of electric and magnetic multipoles $E_{l \pm}$ and $M_{l\pm}$, respectively, where $l = 0, 1, 2, ...$ is the orbital angular momentum of the final-state hadrons and $j=l \pm \frac{1}{2}$ is the total angular momentum. Each multipole is a complex function of energy, which makes the helicity amplitudes complex functions of both energy and scattering angle:

	\begin{subequations}
	\begin{multline}
	\label{eq:hnfull}
	\HN = \sqrt{\frac{1}{2}} \cos\left(\frac{\theta}{2}\right) \sum_{l=0}^{\infty}\left[
	\left(l+2\right)E_{l+}
	+lM_{l+} \right.  \\
	+lE_{\left(l+1\right)-}
	-\left(l+2\right)M_{\left(l+1\right)-}\left. \right]\left(P_l^{'} -P_{l+1}^{'}\right) ,
	\end{multline}

	\begin{multline}
	\label{eq:hspfull}
	\HSP= \sqrt{\frac{1}{2}} \cos\left(\frac{\theta}{2}\right) \sin(\theta) \sum_{l=1}^{\infty}\left[
	E_{l+}-M_{l+} \right]\\
	-E_{\left(l+1\right)-}
	-M_{\left(l+1\right)-}\left. \right]\left(P_l^{''}-P_{l+1}^{''}\right),
	\end{multline}
	
	\begin{multline}	
	\label{eq:hsafull}
	\HSA = \sqrt{\frac{1}{2}} \sin\left(\frac{\theta}{2}\right) \sum_{l=0}^{\infty}\left[
	\left(l+2\right)E_{l+}\right.
	+lM_{l+} \\
	-lE_{\left(l+1\right)-}
	+\left(l+2\right)M_{\left(l+1\right)-}\left. \right]\left(P_l^{'} +P_{l+1}^{'} \right), 
	\end{multline}
	\begin{multline}	
	\label{eq:hdfull}
	\HD = \sqrt{\frac{1}{2}} \sin\left(\frac{\theta}{2}\right) \sin(\theta) \sum_{l=1}^{\infty}\left[
	E_{l+}-M_{l+} \right. \\
	+E_{\left(l+1\right)-}
	+M_{\left(l+1\right)-}\left. \right]\left(P_l^{''}+P_{l+1}^{''} \right). 
	\end{multline}	

	\label{eq:Helicity_Amplitudes}
\end{subequations}
The naming convention for the four helicity amplitudes in Eq. \eqref{eq:Helicity_Amplitudes} follows that of the SAID group \cite{SAID1990}. All 16 single- and double-polarization observables can be written in terms of these four helicity amplitudes; however, in the literature, different sign conventions are used in their definitions. The signs used for the observables in this work are listed in Table \ref{tbl:Observables} and also follow the conventions of the SAID group. 

\begin{table}
	\renewcommand{\arraystretch}{1.5}
	\begin{ruledtabular} 
	\begin{tabular}{c >{ \arraybackslash}p{1.8cm}}
		$ \sigma(\theta) = \frac{q}{2k}\left[|\HN|^2 + |\HD|^2 + |\HSA|^2 + |\HSP|^2\right]$  & \\
		$ \Sigma \: \sigma(\theta) = \qk \text{Re}\left[\HSP\HSAC-\HN\HDC\right]$ & $B$\\
		$T  \: \sigma\left(\theta\right) = \qk \text{Im}\left[\HSP \HNC + \HD \HSAC\right]$&$T$ \\
		$P  \: \sigma\left(\theta\right) = - \qk \text{Im}\left[\HSP \HDC + \HN \HSAC\right]$ & $R$\\  \hline
		$G  \: \sigma\left(\theta\right) = -\qk \text{Im}\left[\HSP \HSAC + \HN \HDC\right]$& $B$, $T$ \\
		$H  \: \sigma\left(\theta\right) = -\qk \text{Im}\left[\HSP \HDC + \HSA \HNC\right]$& $B$, $T$ \\
		$F  \: \sigma\left(\theta\right) = \qk \text{Re} \left[\HSA\HDC + \HSP\HNC\right]$& $B$, $T$\\
		$E  \: \sigma\left(\theta\right) = \frac{q}{2k} \left[|\HN|^2 + |\HSA|^2 - |\HD|^2 - |\HSP|^2\right]$& $B$, $T$\\ \hline
		$O_x  \: \sigma\left(\theta\right) = -\qk \text{Im} \left[\HSA\HDC + \HSP\HNC\right]$& $B$, $R$\\
		$O_z  \: \sigma\left(\theta\right) = -\qk \text{Im} \left[\HSA\HSPC + \HN\HDC\right]$& $B$, $R$\\
		$C_x  \: \sigma\left(\theta\right) = -\qk \text{Re} \left[\HSA\HNC + \HSP\HDC\right]$& $B$, $R$\\
		$C_z \: \sigma\left(\theta\right) = \frac{q}{2k} \left[ |\HSA|^2 + |\HD|^2 - |\HN|^2 - |\HSP|^2\right]$ & $B$, $R$\\ \hline
		$T_x \:  \sigma\left(\theta\right) = \qk \text{Re} \left[\HSP\HSAC + \HN\HDC\right]$& $T$, $R$\\
		$T_z \:  \sigma\left(\theta\right) = \qk \text{Re} \left[\HSP\HNC - \HSA\HDC\right]$& $T$, $R$\\
		$L_x  \: \sigma\left(\theta\right) = \qk \text{Re} \left[\HSA\HNC - \HSP\HDC\right]$& $T$, $R$\\
		$L_z  \: \sigma\left(\theta\right) = \frac{q}{2k} \left[ |\HSP|^2 + |\HSA|^2 - |\HN|^2 - |\HD|^2\right]$& $T$, $R$\\ 
	\end{tabular}
	\caption{List of single-polarization and double-polarization observables analyzed in this work. See Refs. \cite{BDS1, Sandorfi} for a detailed description of the necessary experimental setup and equations for all 16 observables. In the second column, $B$, $T$, and $R$ refer to a measurement of the beam, target, and recoil nucleon polarization, respectively. Note that $\sigma\left(\theta\right) = d\sigma/d\Omega$ is the differential cross section. }
	\label{tbl:Observables}
	\end{ruledtabular}
\end{table}

The literature also mentions measurements of ${d\sigma}/{d\Omega}_{\frac{1}{2}}$ and ${d\sigma}/{d\Omega}_{\frac{3}{2}}$, which are the helicity-dependent cross sections. These observables are linear combinations of the differential cross section and $E$ observables.

\section{Procedure}
Before the analysis could be initiated, some data sets needed modification from their published form. $C_x$ and $C_z$ data from Bradford \textit{et al.}~\cite{Bradford06} were rotated by an angle of $\pi + \theta_{cm}$ from their published form. Here, $\theta_{cm}$ is the polar angle of the outgoing $K^+$ meson in the center-of-momentum frame. (The inverse rotation is given in Eq.~2 in Lleres \textit{et al.} \cite{Lleres09}). When comparing this work with other works, this rotation may also need to be combined with a sign change due to different observable conventions. A ``Rosetta Stone'' that describes different conventions is discussed in Sandorfi \textit{et al.} \cite{Sandorfi2}. This work follows the conventions of the SAID/GWU group for all 16 polarization observables.) $O_x$ and $O_z$ data obtained from Paterson \textit{et al.}~\cite{Paterson16} required the same rotation. After rotation, the Paterson data agree with a previous measurement by Lleres \textit{et al.}~\cite{Lleres09}. The $L_z$ data from Casey~\cite{CaseyPhD} needed to be multiplied by $-1$ due to the $z'$ axis being flipped and a difference in sign conventions between groups. SAPHIR $d\sigma/d\Omega$ measurements from Tran \textit{et al.}\ ~\cite{Tran98} and Glander \textit{et al.}\ ~\cite{Glander04} were removed, as well as a dataset from Hicks \textit{et al.}\ ~\cite{Hicks07}. Finally, data below 1639 MeV from Jude \textit{et al.}\ ~\cite{Jude14} were also removed. 

To begin the analysis, all data were binned into specified small c.m.\ energy ranges. Observables within a single bin were then approximated as functions of just the scattering angle. An appropriate bin width was determined by initially binning the observables into wide bins of 100~MeV and noticing that there was little variation in the double-polarization data over the energy range of the analysis. This meant that 20-MeV wide bins were sufficient to describe the energy variation in the observables near threshold where the $S_{11}$ and $P_{13}$ amplitudes dominate due to the $S_{11}$(1650) and $P_{13}$(1720) resonances. At c.m.\ energies from 1850 to 2200~MeV where resonances are expected to have wider widths, 40~MeV bins were used. 

Once the data were binned, an initial round of single-energy fits was performed in which the $S_{11}$ amplitude was kept real to determine its magnitude. Then a multichannel energy-dependent fit, similiar to those of Shrestha and Manley \cite{ShresthaED}, was performed to determine the $S_{11}$ phase through unitarity constraints. With the $S_{11}$ amplitude fully determined, initial values for the higher-order amplitudes were then determined. An iterative procedure in which we first carried out single-energy fits, followed by a set of multichannel energy-dependent fits of individual partial waves, was used to obtain convergence of the solution to a global minimum.

Because not all measured observables are available in all energy bins, $\chi^2$ penalty terms were added to the standard $\chi^2$ term to constrain the single-energy solutions. The explicit form for a penalty term was
\begin{equation}
\chi^2_{\text{penalty}} = f [(PW^R_{\text{ED}} - PW^R_{\text{fit}})^2 + f^I(PW^I_{\text{ED}} - PW^I_{\text{fit}})^2],
\end{equation}
where $PW^R_{\text{ED}}$ and $PW^I_{\text{ED}}$ are the real and imaginary parts of the partial-wave amplitude found in the preceding energy-dependent fit and $PW^R_{\text{fit}}$ and $PW^I_{\text{fit}}$ are the corrsponding real and imaginary parts of the amplitude determined during each step of the single-energy fit. The factor $f$ was a parameter chosen to control the strength of the penalty term. For the initial round of single-energy fits, we set $f=0$ for no penalty term at all. After the first round of energy-dependent fits, we used values from the energy-dependent fits to constrain selected partial waves in the next round of single-energy fits. This was initially done with a weak penalty constraint (e.g., $f=10$), but as iterations progressed and the energy-dependent solutions did a better job of describing the fitted observables, we gradually increased the strength of the penalty term (e.g., to $f=30$ or $f=50$). This biased results to single-energy solutions that were somewhat similar to the current energy-dependent solution. To minimize bias from the penalty terms, multiple starting solutions were used to determine which produced the best fit. During the analysis, the $\chi^2$ penalty contribution typically remained below a few percent of the total $\chi^2$. 

Once the constrained single-energy solutions converged to agree with the final energy-dependent solution, final error bars on the single-energy amplitudes were obtained by performing ``zero-iteration'' fits in which the phases of the amplitudes were held fixed and only their moduli were allowed to vary.  During this step, the penalty terms were removed from the analysis.  

\section{Results}
This section presents final results for the partial-wave analysis of \GPKL \ and predictions for the reaction \GNKL. It compares our results with those of the BnGa group and shows the quality of agreement with integrated cross-section data , which were not directly fitted. Table \ref{tab:BnGavsKSU2} shows the $\chi^2$ breakdown by observable and compares our results with the BnGa 2016 solution.  The table also lists references for each observable included in the single-energy fits.

The resonances that were found to contribute significantly in our multichannel energy-dependent fits were the $S_{11}$(1650), $P_{11}$(1880), $P_{13}$(1720), $P_{13}$(1900), $D_{13}$(2120), and $D_{15}$(2080). This is similar to the resonance structure found by other groups. There was some indication in the data of a possible $F_{17}$ resonance around 2300~MeV that was also seen in pion photoproduction. This is in agreement with quark-model predictions \cite{Capstick98} that a higher-lying $F_{17}$ resonance should couple to $K\Lambda$.

		\begin{table*}[htpb]
			\centering
			\renewcommand{\arraystretch}{1.5}
			\begin{ruledtabular} 
			\begin{tabular}{ccccc}

				Observable	& KSU & BnGa (2016) & No.\ Data & References \\   \hline
				\DSG    	& 8400 & 9500 & 4101 & \cite{Donoho58,Brody60,Anderson62,Peck64,Anderson65,Mori66,Groom67,Bleckmann70,Decamp70,Fujii70,Goeing71,Feller72,Bockhorst94,Mcnabb04,Bradford06,Sumihama06,Mccracken10,Jude14} \\
				$T$	        & 3000 & 2100 & 451  & \cite{Althoff78,Lleres09,WolfordPhD,Paterson16} \\
				$\Sigma$        & 2400 & 1200 & 418  & \cite{Zegers03,Sumihama06,Lleres07,Paterson16} \\
				$P$     	& 3200 & 3200 & 1565 & \cite{Mcdaniel59,Thom63,Borgia64,Grilli65,Groom67,Fujii70,Haas78,Bockhorst94,Mcnabb04,Lleres07,Mccracken10,Paterson16} \\
				$F$     	& 990 & 5300 & (84)  & \cite{WolfordPhD} \\
				$E$     	& 2500 & 330 &  (72) & \cite{CaseyPhD} \\
				$C_x$	& 310 & 230 &  97    & \cite{Bradford07} \\
				$C_z$	& 430 & 210 &  97    & \cite{Bradford07} \\
				$O_x$	& 1000 & 460 &  363  & \cite{Lleres09,Paterson16} \\
				$O_z$	& 1500 & 650 &  363  & \cite{Lleres09,Paterson16} \\
				$T_x$	& 1000 & 4200 & (77) & \cite{WolfordPhD} \\
				$T_z$	& 1300 & 2700 & (53) & \cite{WolfordPhD} \\
				$L_x$	& 90 & 95 &  (87)    & \cite{CaseyPhD} \\
				$L_z$	& 230 & 180 & (72)   & \cite{CaseyPhD} \\
				\hline
				Fit Total& 26000 & 30000  & \\

			\end{tabular}
			
			\caption{Comparison of $\chi^2$ contributions to $\gamma p \to K^+ \Lambda$ for different observables and analyses. Column one lists the observable, columns two and three list the $\chi^2$ contribution from each observable, column four lists the number of published (unpublished) data points for the observable, and column five lists references for the observable. See text for information on binning changes and discussion on the points included. The c.m.\ energy range and binning used to generate the $\chi^2$ values was $W = 1610$ - 2200~MeV in 10-MeV wide bins.}
			\label{tab:BnGavsKSU2}	
			\end{ruledtabular} 				
		\end{table*}

\begin{figure}[ht]
	\centering
	\includegraphics[scale=.40,trim=0mm 11mm 5mm 18mm,clip=true]{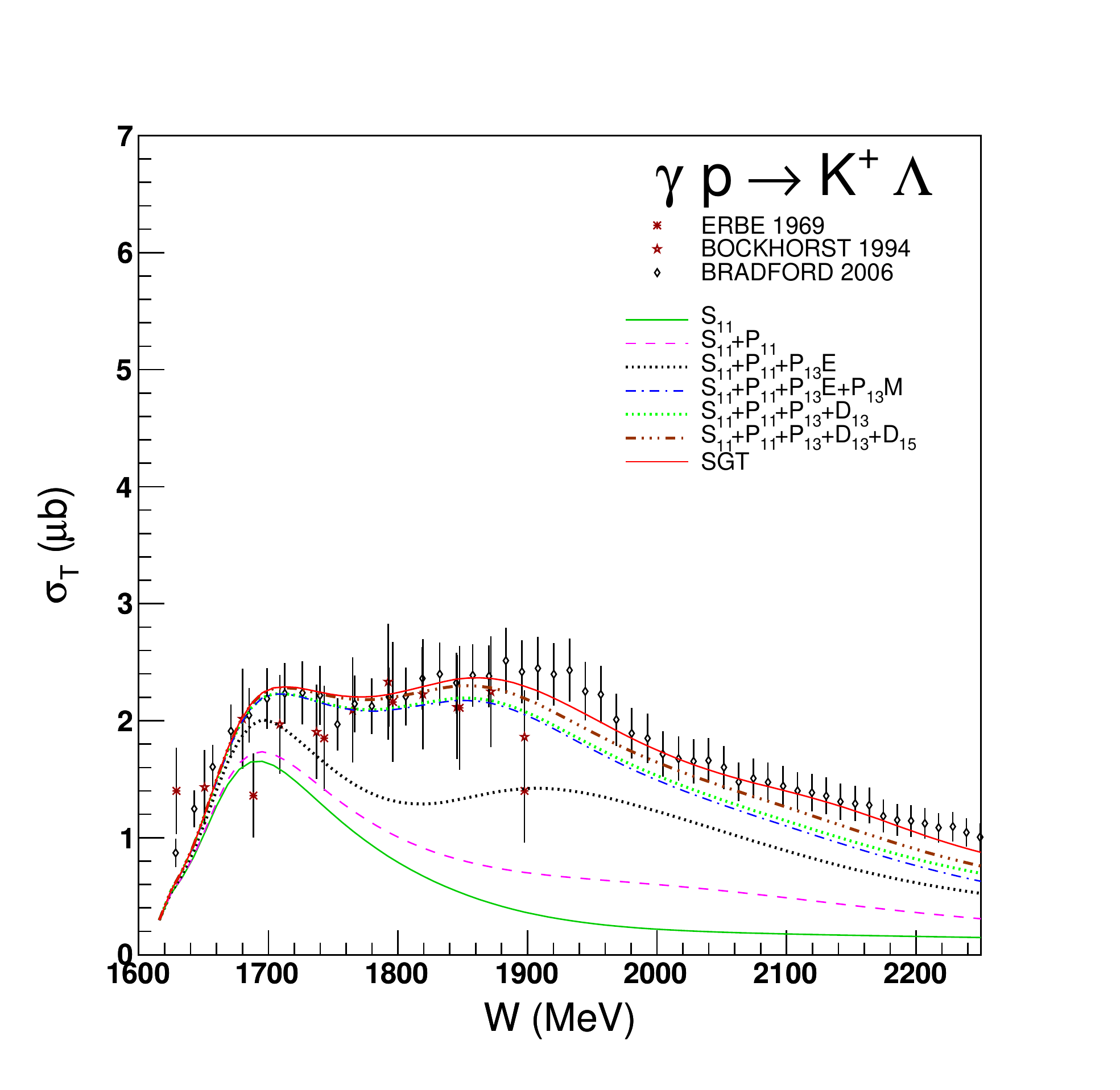}
	\caption[Integrated cross section for \GPKL.]{\label{SGTGPKL} Integrated cross section for \GPKL. The curves show the contribution to the cross section by successively adding each partial wave. Points are from ERBE 1969 \cite{Erbe69}, BOCKHORST 1994 \cite{Bockhorst94}, and BRADFORD 2006 \cite{Bradford06}.  }
	
\end{figure}

\begin{figure}[tph]
	\centering
	\includegraphics[scale=.40,trim=0mm 10mm 5mm 22mm,clip=true]{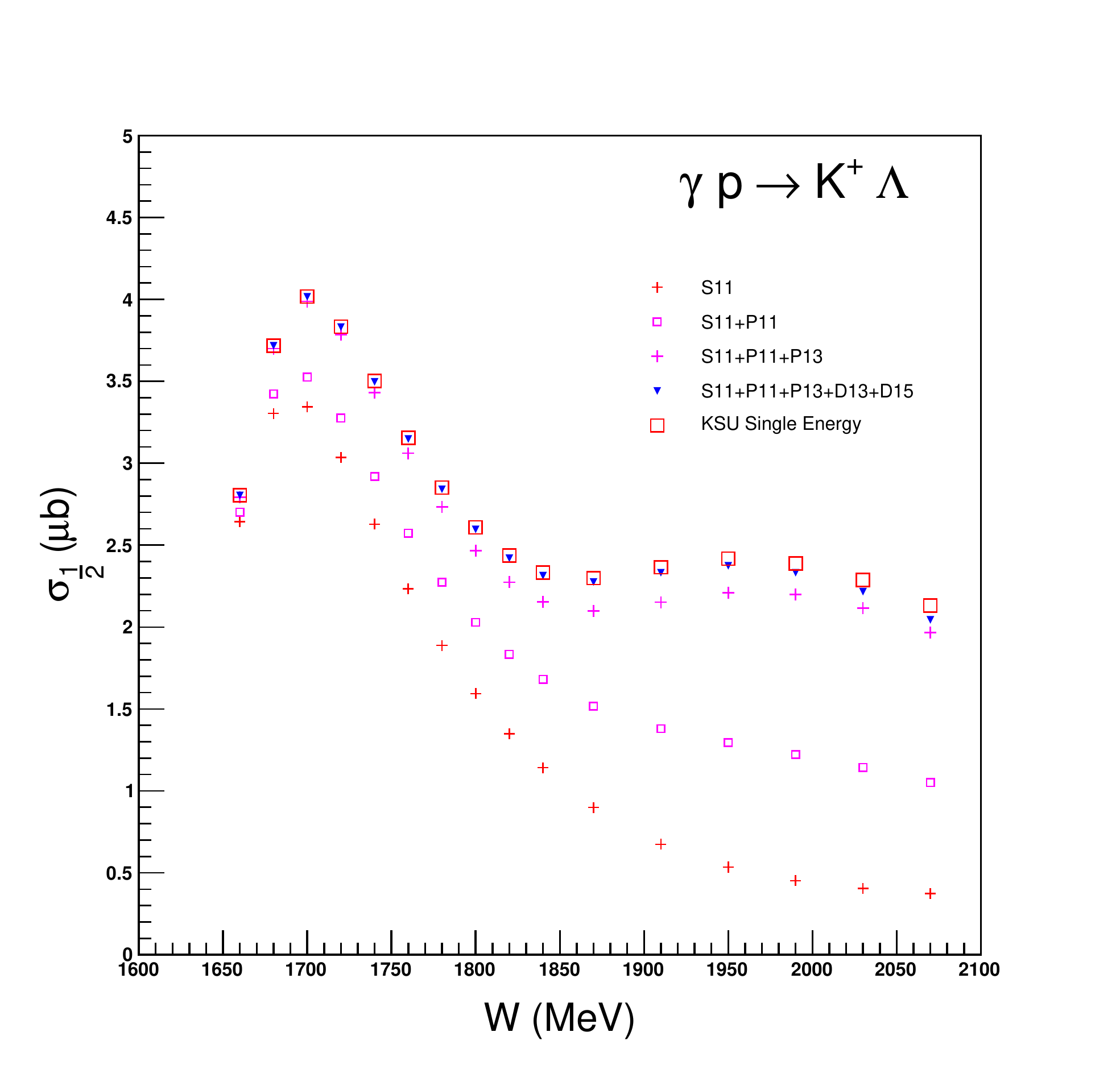}
	\caption{Prediction of helicity-$1/2$ integrated cross section for \GPKL. The plot shows the predicted contribution to the cross section by successively adding each partial wave.}
	\label{SGTGPKL12}
\end{figure}

\begin{figure}[ht]
	\centering
	\includegraphics[scale=.99,trim=20mm 194mm 110mm 18mm,clip=true]{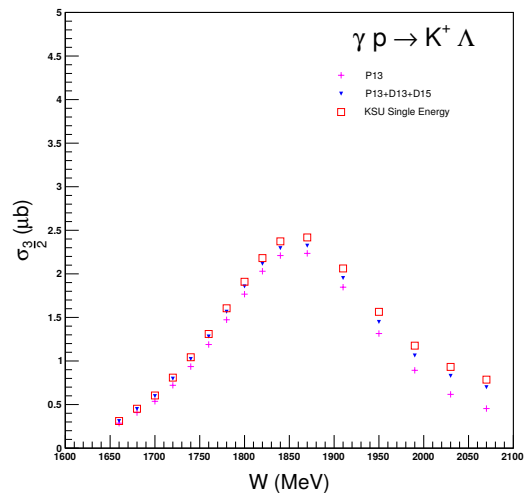}
	\caption{Prediction of helicity-$3/2$ integrated cross section for \GPKL. The plot shows the predicted contribution to the cross section by successively adding each partial wave.}
	\label{SGTGPKL32}
\end{figure}

The integrated cross section for \GPKL \ is dominated by the $S_{11}$ amplitude at low energies and the $P_{13}$ amplitude at higher energies. The cross section is then saturated by small contributions from the other partial waves up to and including $G_{19}$, which was the highest partial wave included in our fits of $\gamma p \rightarrow K^+ \Lambda$ data. There is excellent agreement between the results of the analysis and the data. Figures \ref{SGTGPKL}, \ref{SGTGPKL12}, and \ref{SGTGPKL32} show the integrated cross section as well as predictions of the helicity-1/2 and -3/2 integrated cross sections.  The helicity-3/2 cross section is predicted to be strongly dominated by the $P_{13}$ multipoles.

\begin{figure*}
	\centering
	\includegraphics[scale=0.99,trim={20mm} {43mm} {\trimCC} {\trimDD},clip=true]{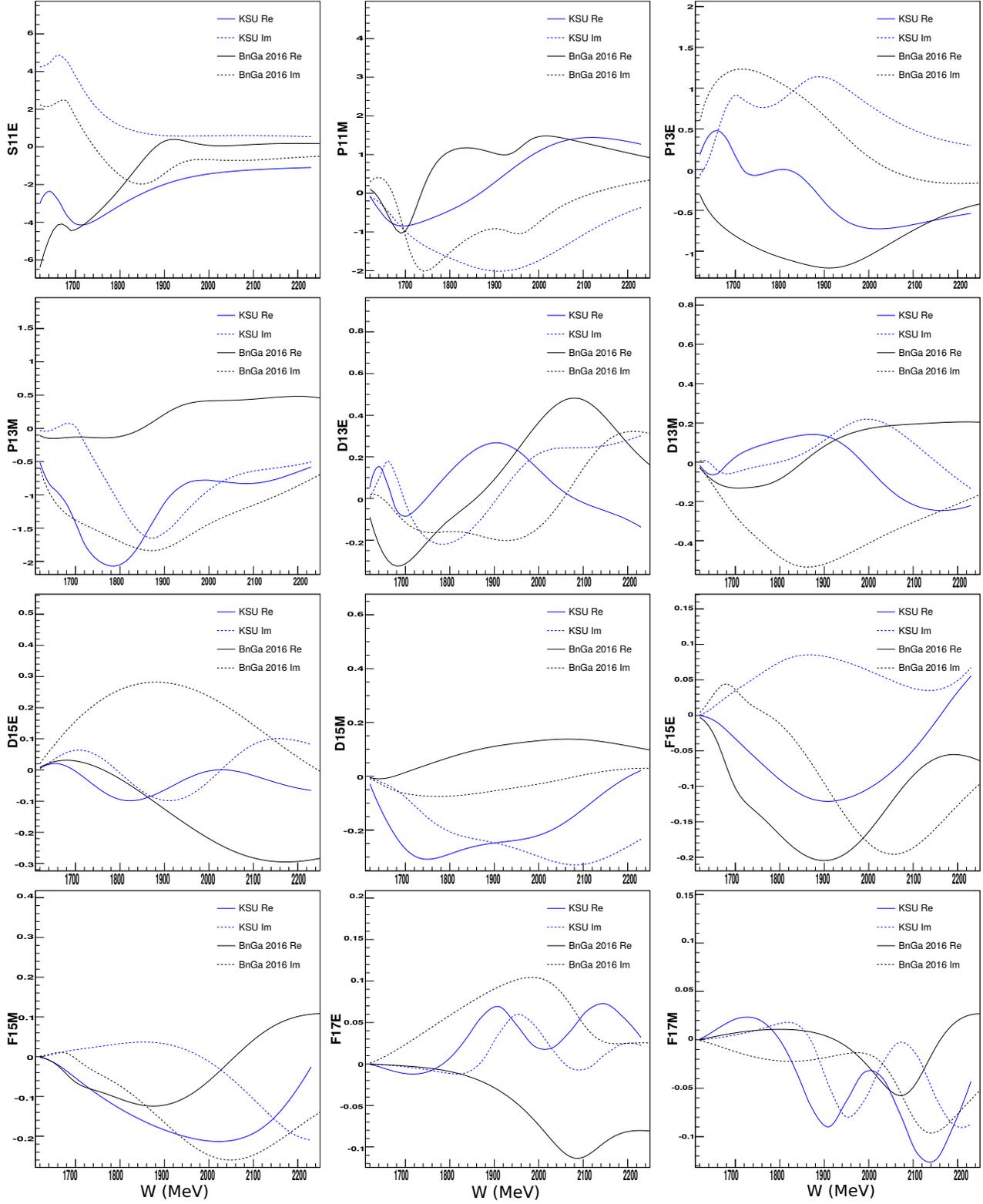}		
	\caption{Comparison of the individual \GPKL \ partial-wave amplitudes for each group. Blue curves are from this work and black curves are from BnGa 2016. The solid and dotted curves are, respectively, the real and imaginary parts of the amplitudes, which are in units of mfm.}	
	\label{GPKL_PWA_Compare1}
\end{figure*}

Plots comparing the partial-wave amplitudes for this reaction are shown in Fig. \ref{GPKL_PWA_Compare1}, with comparisons only available between this work and BnGa 2016 \cite{Sarantsev}. 

$S_{11}$ is the only amplitude from this work that agrees with results from the BnGa group. This was unexpected because of the sizable number of spin observables that have been measured for $\gamma p \to K^+ \Lambda$. However, not all discrepancies are large. For instance, differences in the $P_{11}M$ amplitude seem to be correlated with a different mass and width for its resonance parameters as the behavior in the two amplitudes is similar. 

In the $P_{13}E$ amplitude, differences are more significant. Resonance behavior is expected around 1720~MeV based on the photo and $K \Lambda$ couplings. For a resonant amplitude, the real part should approach zero near the resonance and the imaginary part should peak, a behavior in the amplitude that is found in this work. An odd behavior is found in the BnGa results for the amplitude $P_{13}M$. At low energies the amplitude shows a behavior like a Born term (which only contains a real part), but is found in the imaginary part instead. Their real part, also does not show a turn towards zero near the resonance. This suggests that perhaps there may be a global phase problem with the BnGa solution at low energies.

\begin{figure*}
	\centering
	\includegraphics[scale=0.98,trim={21mm} {172mm} {20mm} {12mm},clip=true]{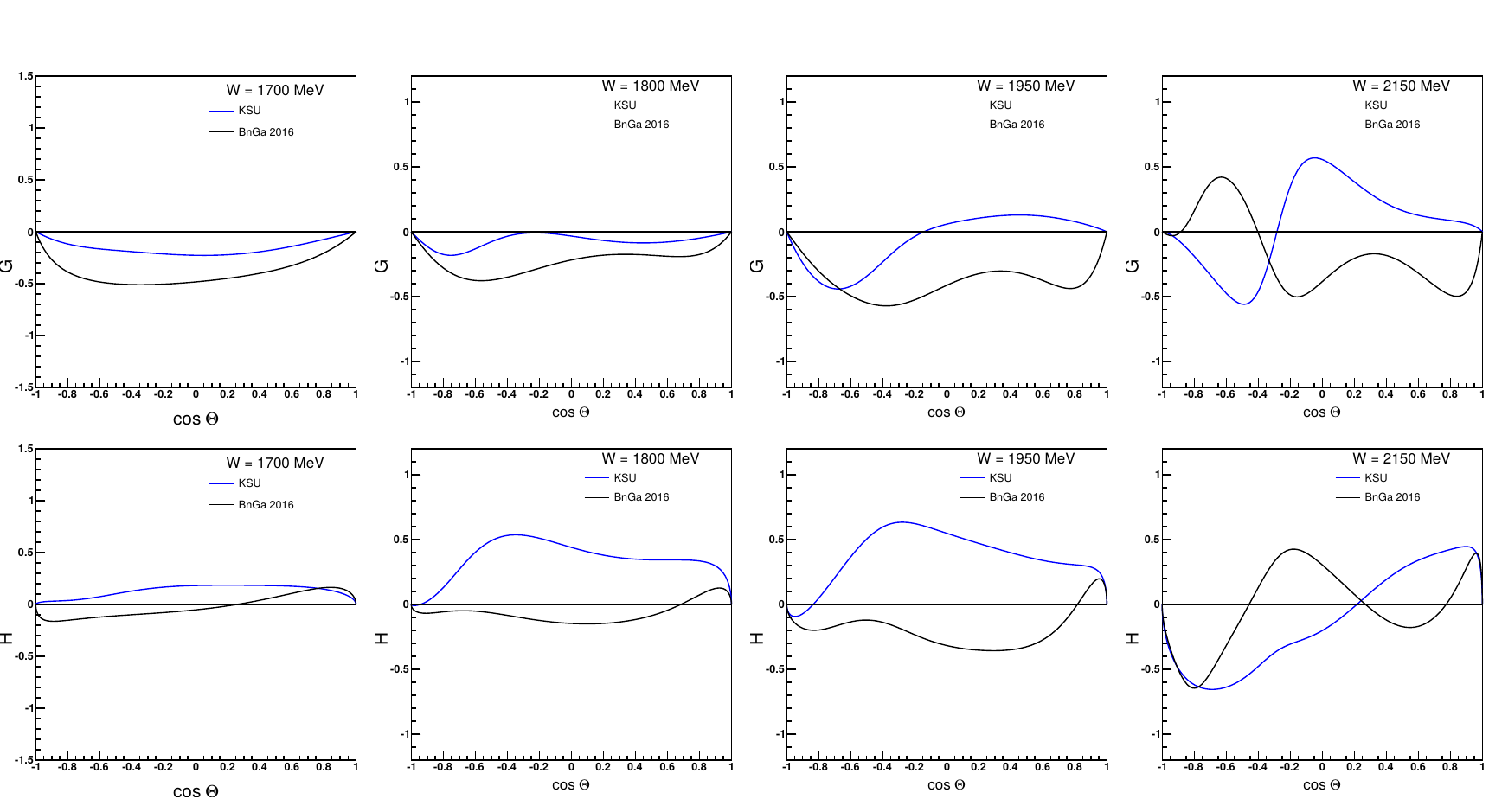}
	\caption{The top four figures show predictions of the observable $G$ and the bottom four figures show predictions of the observable $H$ for \GPKL \ at c.m. energies 1700, 1800, 1950, and 2150 MeV.}	
	\label{GPKL_PredictG}
\end{figure*}

Figure \ref{GPKL_PredictG} shows predictions for the observables $G$ and $H$ for this work and BnGa 2016 at selected energies. Despite the large number of observables that have been measured, predictions of $G$ and $H$ still show significant disagreement over the full energy range. 

After completion of our analysis, we learned that $\gamma p \rightarrow K^+ \Lambda$ has also been studied recently within the J\"ulich-Bonn (J\"uBo) coupled-channel framework \cite{Roenchen2018}. The level of agreement of the new J\"uBo solution with $d\sigma/d\Omega$, $\Sigma$, $P$, $O_x$, $O_z$, $C_{x'}$, and $C_{z'}$ data is quite good overall.  Our solution tends to give a more isotropic differential cross section near threshold, although a detailed comparison of our solutions has not been evaluated.  It is clear, however, that the multipoles in J\"uBo solution do not agree with those of either the BnGa group or with our solution. As noted previously, our $S_{11}$ amplitude qualitatively agrees with the BnGa results, whereas the $S_{11}$ (or $E_{0+}$) amplitude in the new  J\"uBo solution agrees with neither our solution nor with the corresponding BnGa amplitude.  Additional double-polarization data are probably needed to resolve these differences.

\begin{figure*}
	\includegraphics[width=.98 \linewidth,trim=21mm 40mm 18mm 16mm,clip=true]{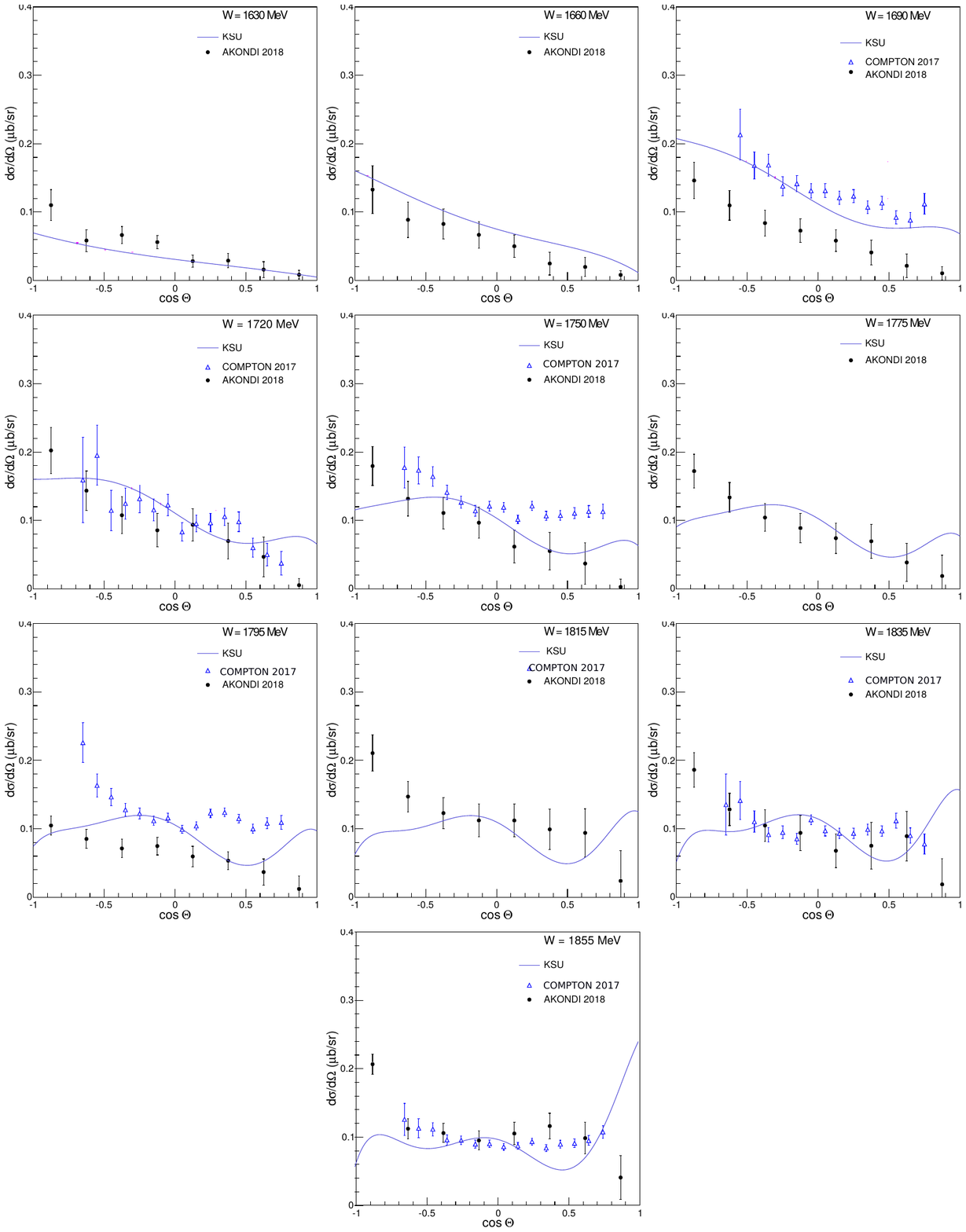}
	\caption{Predictions of $d\sigma/d\Omega$ for $\gamma n \rightarrow K^0 \Lambda$ at $W$ = 1630 to 1855~MeV. Data points are from COMPTON 2017 \cite{Compton2017} and AKONDI 2018 \cite{AKONDIDiss}. }
	\label{PredictGNKL}
\end{figure*}

\begin{figure}
	\includegraphics[scale = .48,trim=18mm 105mm 36mm 15mm,clip=true]{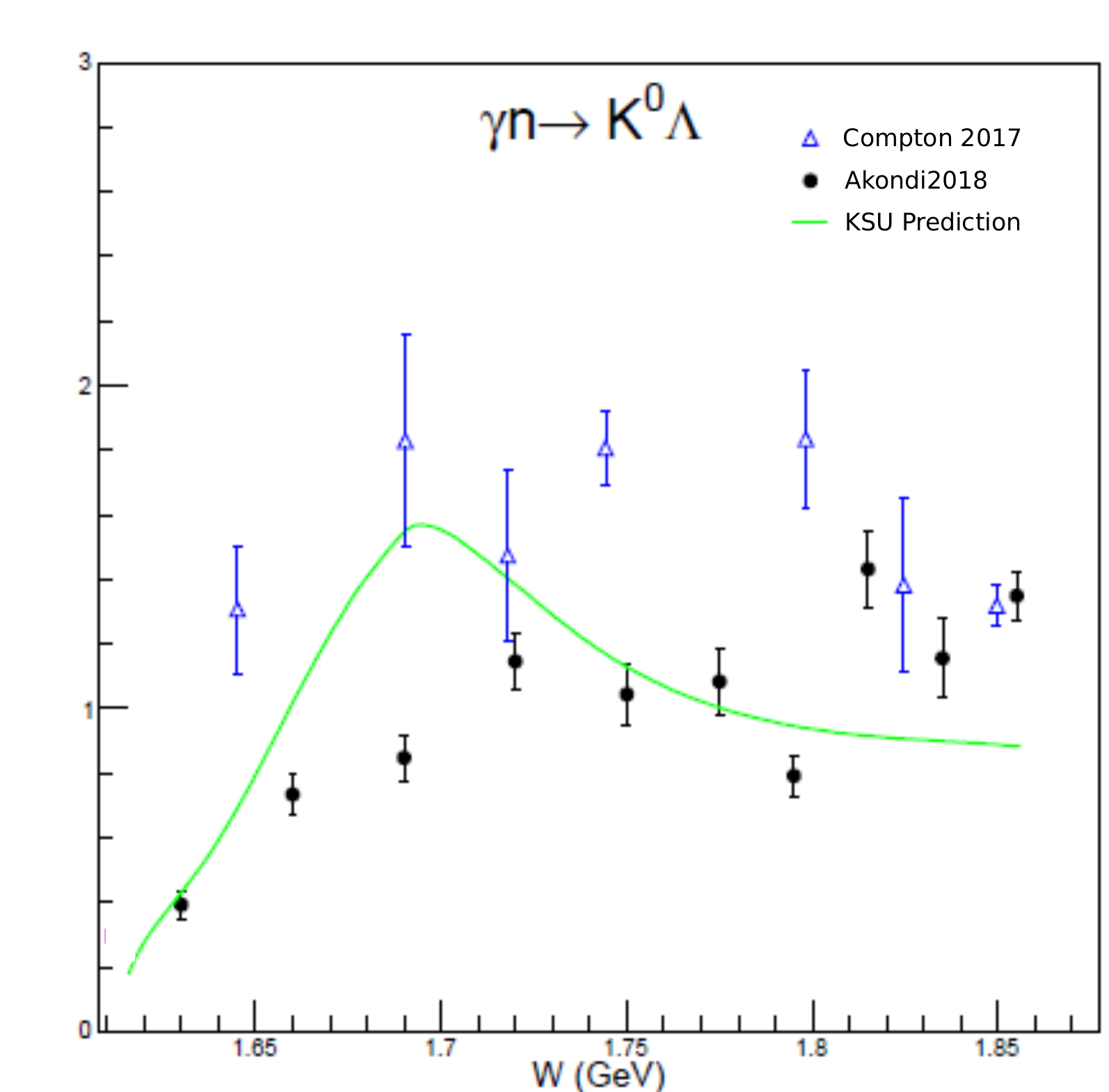}	
	\caption{Predicted integrated cross section for $\gamma n \rightarrow K^0 \Lambda $. The green curve shows our prediction and the data points are from COMPTON 2017 \cite{Compton2017} and AKONDI 2018 \cite{AKONDIDiss}.}
	\label{PredictGNKLSGT}

\end{figure}

Because this analysis was carried out in conjunction with other photoproduction and hadronic reactions that included $\gamma n$ and $K \Lambda$ states, we are able to make predictions for $\gamma n \rightarrow K^0 n$.  The highest partial wave included in our predictions for $\gamma n \rightarrow K^0 \Lambda$ was $F_{15}$ and our predictions are expected to be reasonable up to c.m.\ energies near 1900~MeV. Our predictions are compared to differential cross-section data by CLAS \cite{Compton2017} and Akondi \cite{AKONDIDiss} in Fig.~\ref{PredictGNKL}.  Our prediction for the integrated cross section is shown in Fig.~\ref{PredictGNKLSGT}.
The agreement is reasonably good with the $d\sigma/d\Omega$ CLAS data and Akondi results over most of the angular range below about 1800~MeV while there are a few places above that energy where the prediction has a bump at forward angles not seen in the data. 
The quality of agreement of our prediction with the CLAS \cite{Compton2017} and Akondi \cite{AKONDIDiss} measurements suggests that the couplings to $\gamma n$ and $K \Lambda$ are highly constrained by the other reactions. It also lends credence to the fits presented in this work.

\section{Summary and Conclusions}

This work presents results from a partial-wave analysis of $\gamma p \to K^+ \Lambda$ and predictions for $\gamma n \to K^0 \Lambda$. Results from previous works that $S_{11}$ and $P_{13}$ were the dominant amplitudes contributing to the integrated cross section were confirmed; however, the amplitudes from this work show more resonance-like behavior with less background than found by previous works. A potential second $F_{17}$ resonance near 2300~MeV was seen in this reaction as well as in $\pi N \rightarrow \pi N$ elastic scattering and $\gamma N \to \pi N$ photoproduction. More data above 2300~MeV are necessary to confirm its existence and its properties.

This work suggests that data for more than eight independent observables may be needed to reach a single unique solution for photoproduction reactions due to the large uncertainties in the double-polarization measurements and inconsistencies in the data.

The $\gamma p \to K^+ \Lambda$ amplitudes from this work have been included in an updated multichannel energy-dependent partial-wave analysis \cite{paper3} that also incorporates our single-energy amplitudes for $\gamma p \rightarrow \eta p$  and $\gamma n \rightarrow \eta n$ \cite{paper1}.  In Ref.~\cite{paper3}, we present and discuss the resonance parameters obtained from a fit of single-energy amplitudes for these reactions combined with corresponding amplitudes for $\gamma N \rightarrow \pi N$, $\pi N \rightarrow \pi N$, $\pi N \rightarrow \pi \pi N$, $\pi N \rightarrow K \Lambda$, and $\pi N \rightarrow \eta N$.  Reference~\cite{paper3} also includes Argand diagrams that compare the results of our single-energy fits with our final energy-dependent partial-wave amplitudes.

\begin{acknowledgments}

The authors would like to thank Professor Igor Strakovsky for supplying much of the database and Dr.~C.~S.\ Akondi for providing plots of the $\gamma n \rightarrow K^0 \Lambda$ data as well as access to preliminary data. This work was supported in part by the U.S.\ Department of Energy under Awards No.\ DE-FG02-01ER41194 and DE-SC0014323, and by the Department of Physics at Kent State University. 
\end{acknowledgments}





\appendix
\section{Final Fits to Experimental Data}
Figures \ref{GPKLFirst} - \ref{GPKLLast} compares our energy-dependent solution to the data included in our analysis. The partial-wave amplitudes used to generate the curves are available in the form of data files \cite{Datafile}. Also shown in each figure are the fits from BnGa 2016 \cite{Sarantsev}. 

The sources for the data points found in the figures for this reaction are DONOHO 1958 \cite{Donoho58}, MCDANIEL 1959 \cite{Mcdaniel59}, BRODY 1960 \cite{Brody60},  ANDERSON 1962 \cite{Anderson62}, THOM 1963 \cite{Thom63}, BORGIA 1964 \cite{Borgia64}, PECK 1964 \cite{Peck64},  ANDERSON 1965 \cite{Anderson65}, GRILLI 1965 \cite{Grilli65}, MORI 1966 \cite{Mori66}, GROOM 1967 \cite{Groom67}, BLECKMANN 1970 \cite{Bleckmann70}, DECAMP 1970 \cite{Decamp70}, FUJII 1970 \cite{Fujii70}, GOING 1971 \cite{Goeing71}, FELLER 1972 \cite{Feller72}, ALTHOFF 1978 \cite{Althoff78}, HAAS 1978 \cite{Haas78}, BOCKHORST 1994 \cite{Bockhorst94},  ZEGERS 2003 \cite{Zegers03}, MCNABB 2004 \cite{Mcnabb04}, BRADFORD 2006 \cite{Bradford06}, SUMIHAMA 2006 \cite{Sumihama06}, BRADFORD 2007 \cite{Bradford07}, LLERES 2007 \cite{Lleres07},  LLERES 2009 \cite{Lleres09}, MCCRACKEN 2010 \cite{Mccracken10}, CASEY 2011 \cite{CaseyPhD}, JUDE 2014 \cite{Jude14}, WOLFORD 2014 \cite{WolfordPhD}, and PATERSON 2016 \cite{Paterson16}. 


\onecolumngrid
\renewcommand{\pathGraphing}{\pathGPKL}
\renewcommand{\printReac}{\GPKL}
\renewcommand{\ObsName}{DSG}
\renewcommand{\printObs}{\DSGT}
\newcommand{\scaleFac}{0.99}

\newcommand{\trimA}{18mm}
\newcommand{\trimB}{28mm}
\newcommand{\trimC}{2mm}
\newcommand{\trimD}{14mm}
\newcommand{\figWidth}{.40}

\noindent
\begin{figure}
	\includegraphics[scale=\scaleFac,trim={\trimA} {\trimB} {\trimC} {\trimD},clip=true]{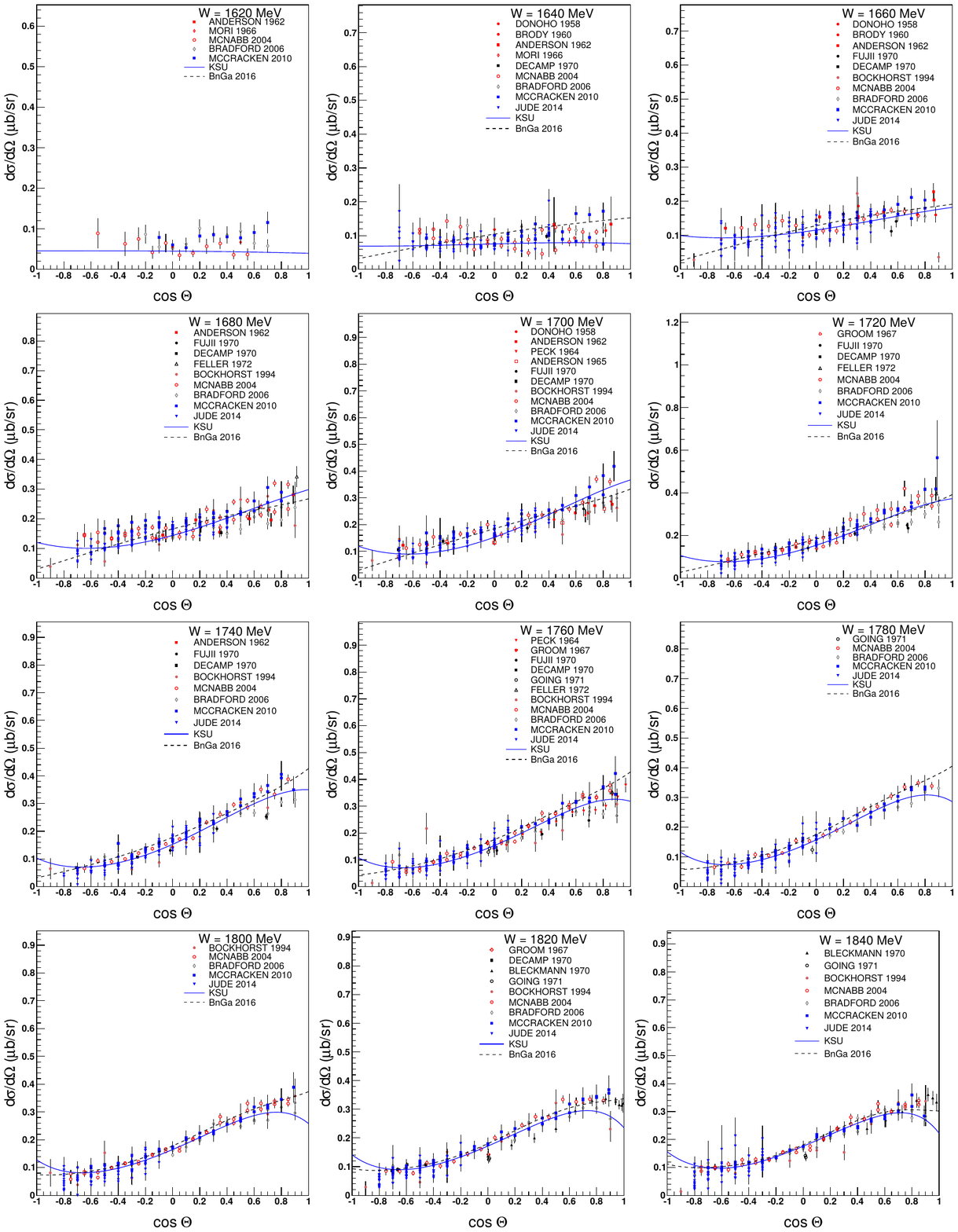}
	\caption{\label{GPKLFirst}Fits to \printObs{} for \printReac{} at $W$ = 1620 to 1840~MeV. See text for references.}
\end{figure}

\noindent
\begin{figure}
	\includegraphics[scale=\scaleFac,trim={\trimA} {\trimB} {\trimC} {\trimD},clip=true]{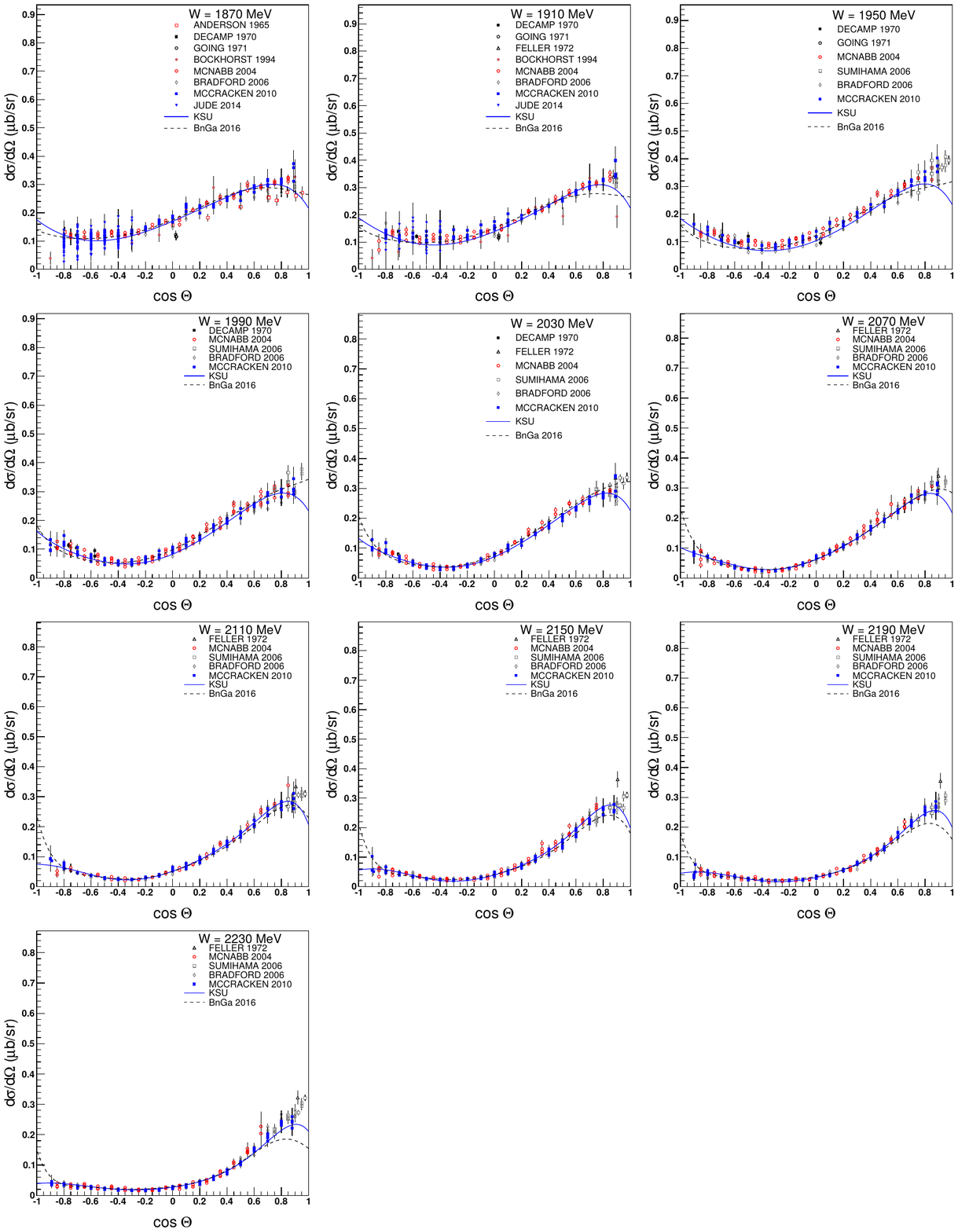}
	\caption{Fits to \printObs{} for  \printReac{} at $W$ = 1870 to 2230~MeV.  See text for references.}
\end{figure}

\renewcommand{\ObsName}{S}
\renewcommand{\printObs}{$\Sigma$}
\noindent
\begin{figure}
	\includegraphics[scale=\scaleFac,trim={\trimA} {\trimB} {\trimC} {\trimD},clip=true]{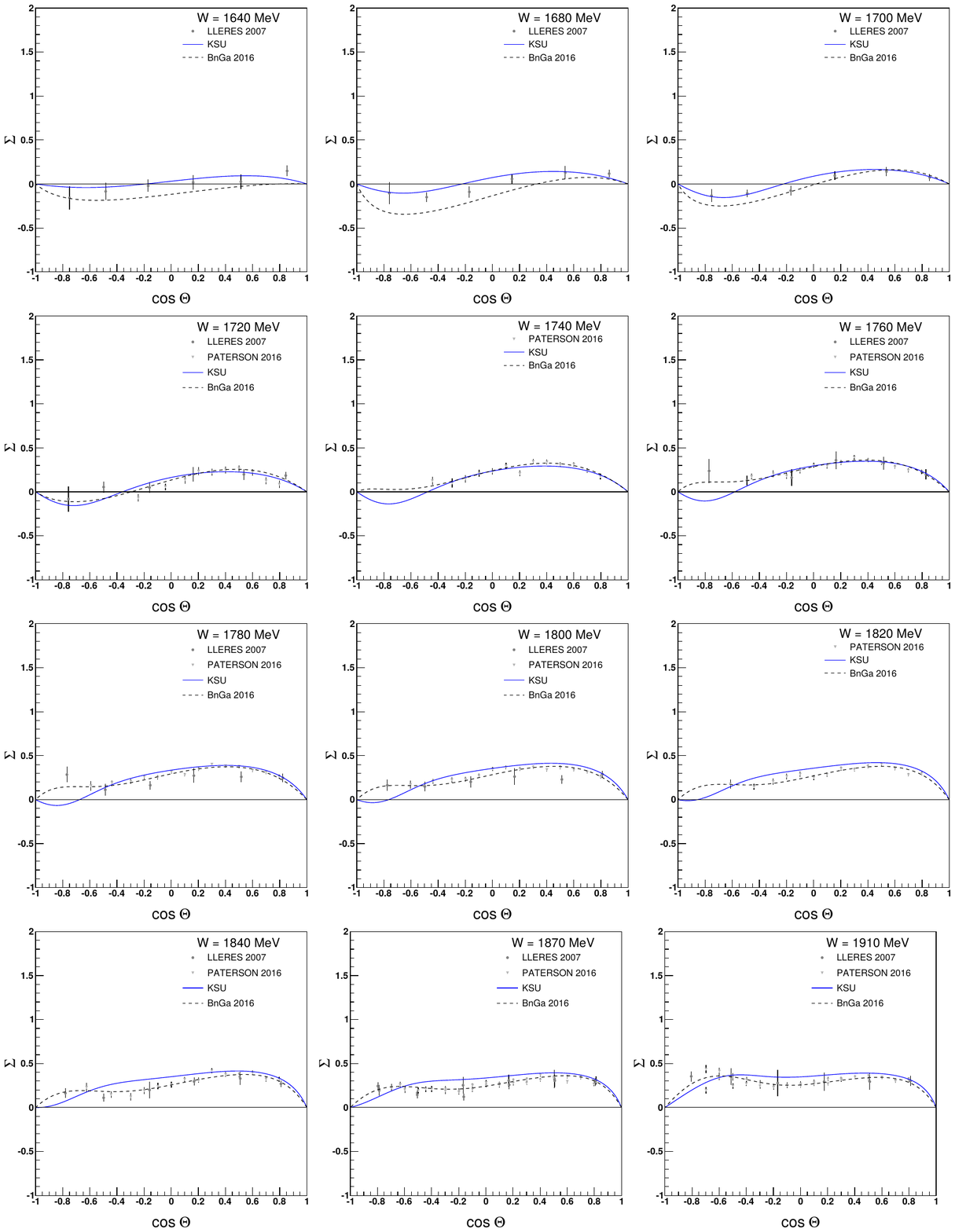}
	\caption{Fits to \printObs{} for \printReac{} at $W$ = 1640 to 1910~MeV.  See text for references.}
\end{figure}

\noindent
\begin{figure}
	\includegraphics[scale=\scaleFac,trim={\trimA} {\trimB} {\trimC} {\trimD},clip=true]{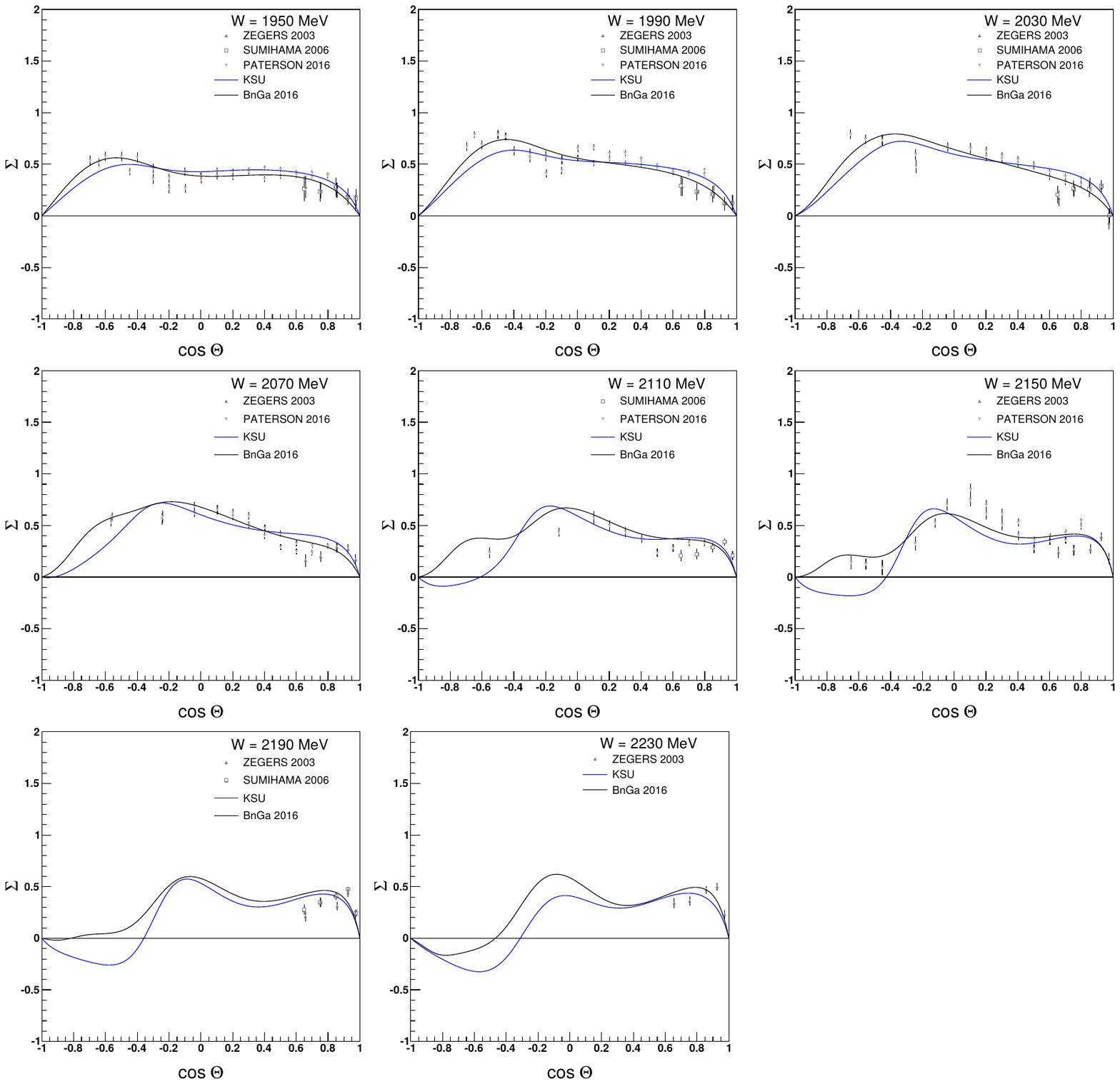}
	\caption{Fits to \printObs{} for \printReac{} at $W$ = 1950 to 2230~MeV.  See text for references.}
\end{figure}

\renewcommand{\ObsName}{T}
\renewcommand{\printObs}{$T$}
\noindent
\begin{figure}
	\includegraphics[scale=\scaleFac,trim={\trimA} {\trimB} {\trimC} {\trimD},clip=true]{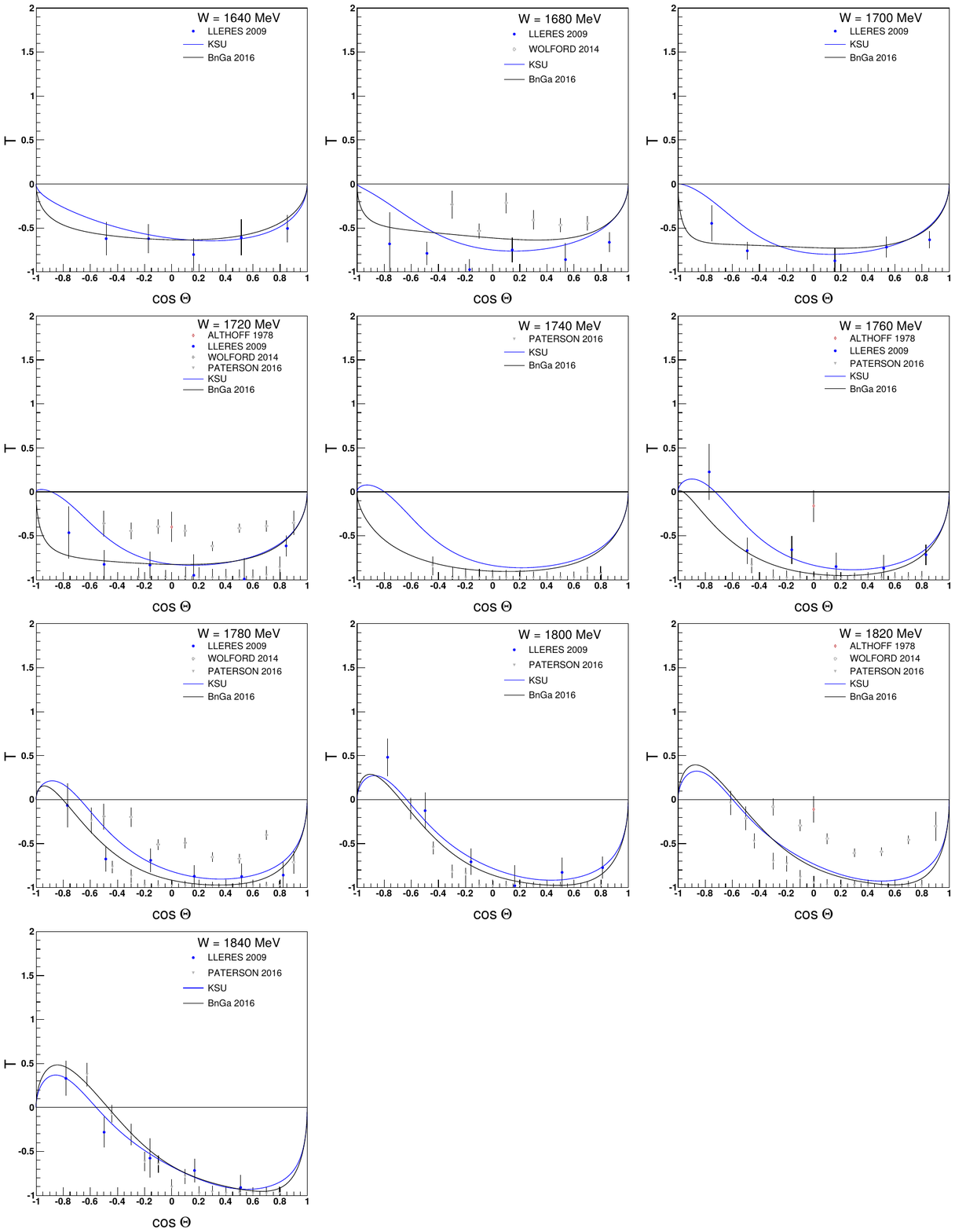}
	\caption{Fits to \printObs{} for \printReac{} at $W$ = 1640 to 1840~MeV.  See text for references.}
\end{figure}

\noindent
\begin{figure}
	\includegraphics[scale=\scaleFac,trim={\trimA} {\trimB} {\trimC} {\trimD},clip=true]{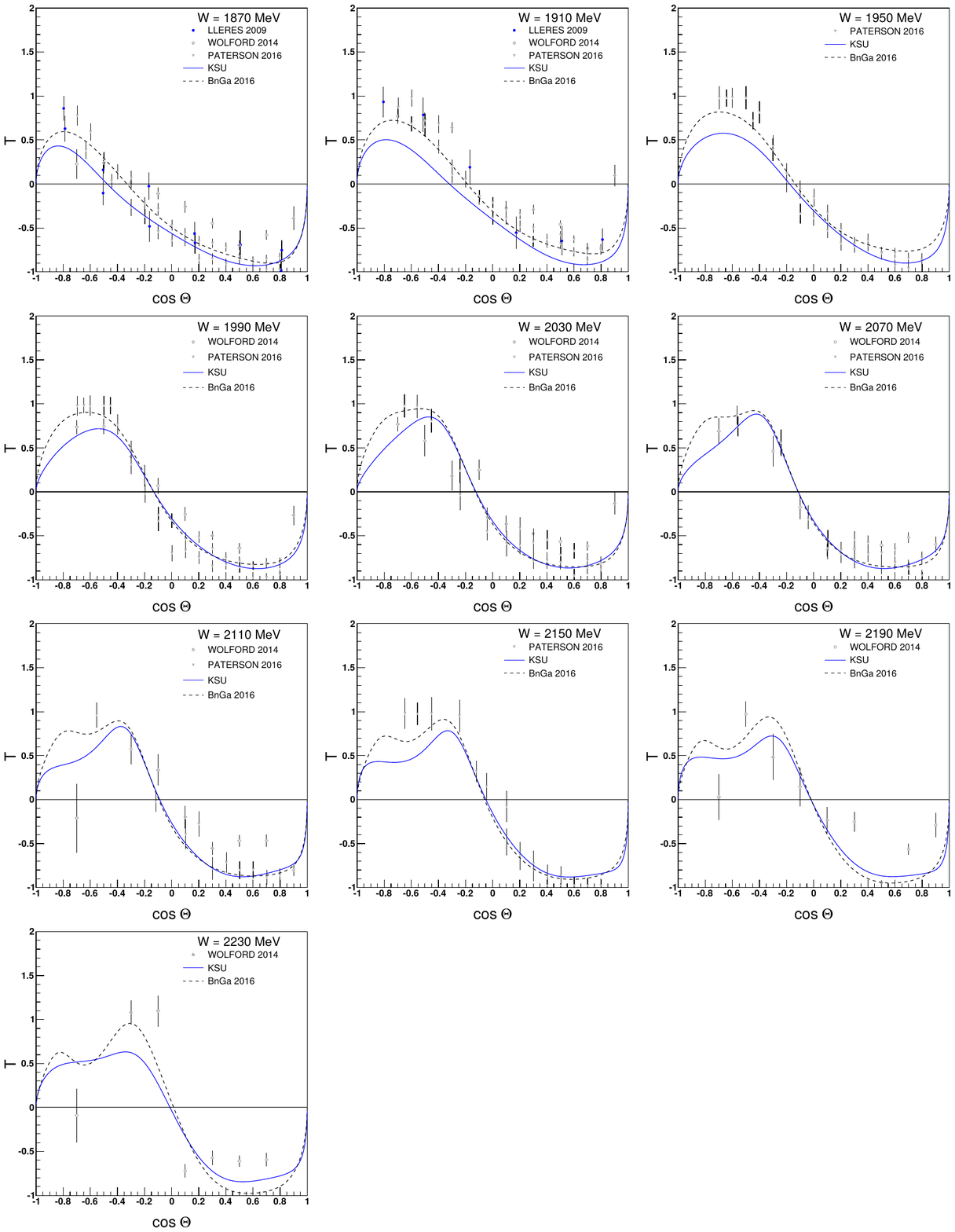}
	\caption{Fits to \printObs{} for \printReac{} at $W$ = 1870 to 2230~MeV.  See text for references.}
\end{figure}

\renewcommand{\ObsName}{P}
\renewcommand{\printObs}{$P$}
\noindent	
\begin{figure}
	\includegraphics[scale=\scaleFac,trim={\trimA} {\trimB} {\trimC} {\trimD},clip=true]{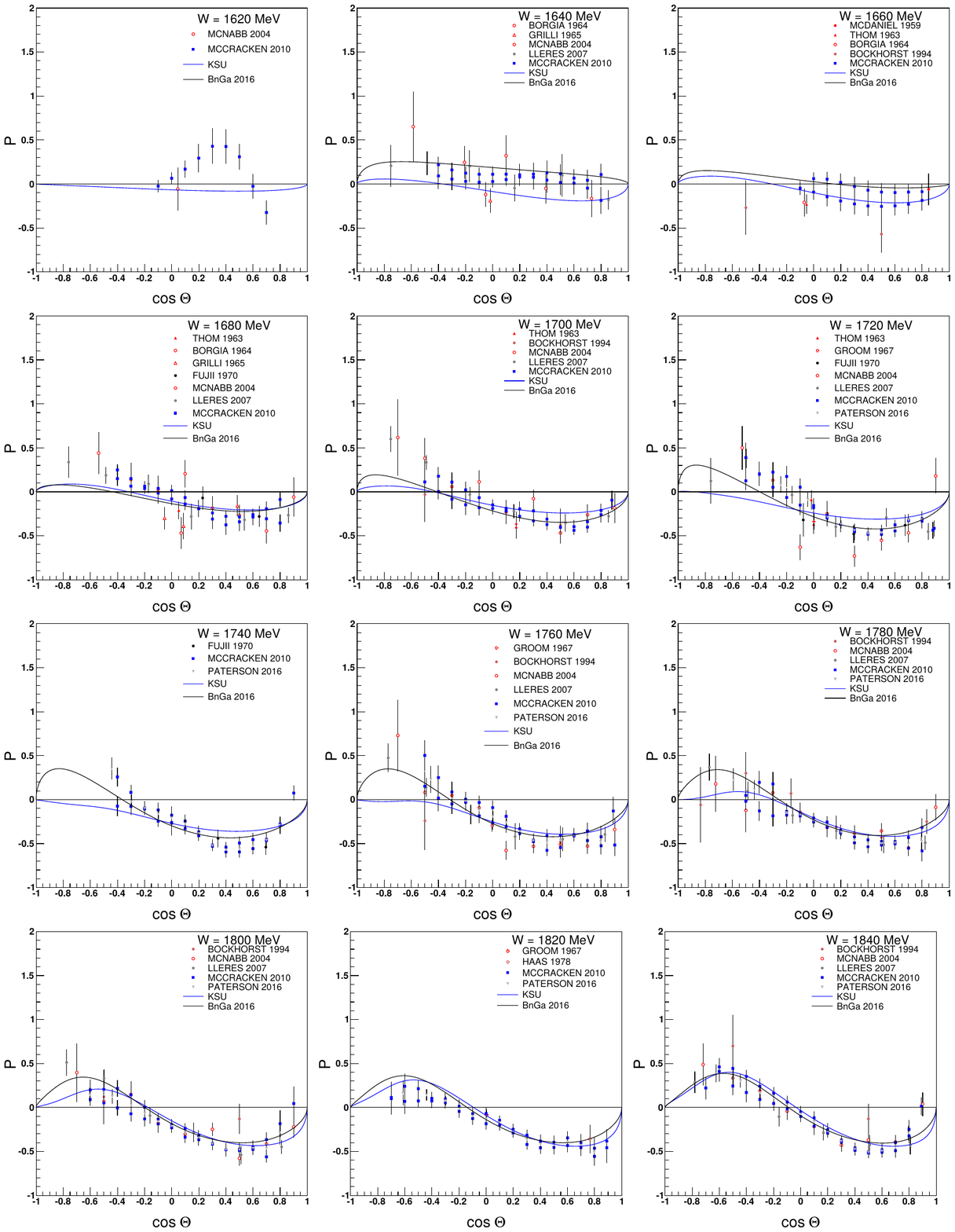}
	\caption{Fits to \printObs{} for \printReac{} at $W$ = 1620 to 1840~MeV.  See text for references.}
\end{figure}

\noindent
\begin{figure}
	\includegraphics[scale=\scaleFac,trim={\trimA} {\trimB} {\trimC} {\trimD},clip=true]{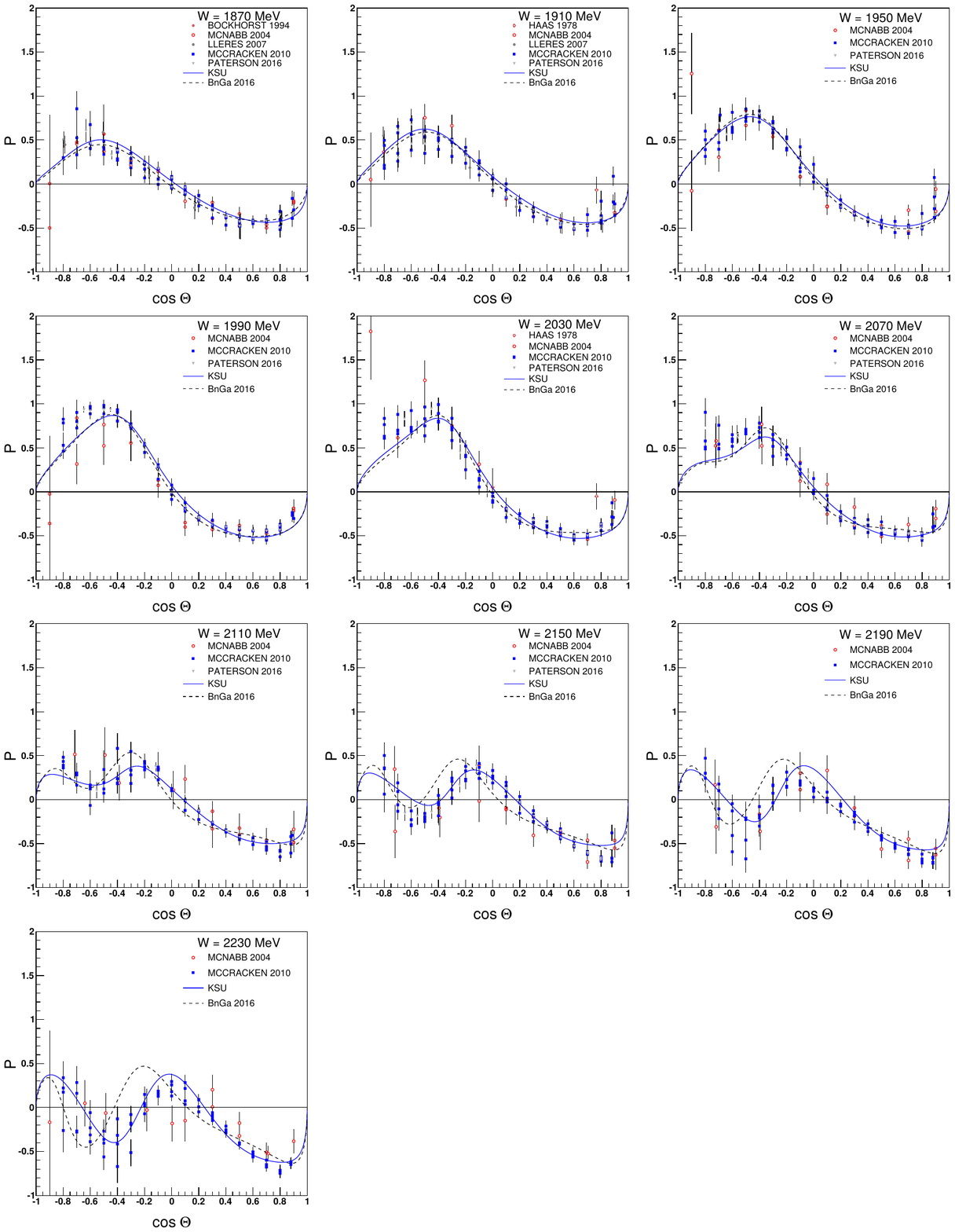}	
	\caption{Fits to \printObs{} for \printReac{} at $W$ = 1870 to 2230~MeV.  See text for references.}
\end{figure}

\renewcommand{\ObsName}{E}
\renewcommand{\printObs}{$E$}
\noindent
\begin{figure}
	\includegraphics[scale=\scaleFac,trim={\trimA} {88mm} {\trimC} {\trimD},clip=true]{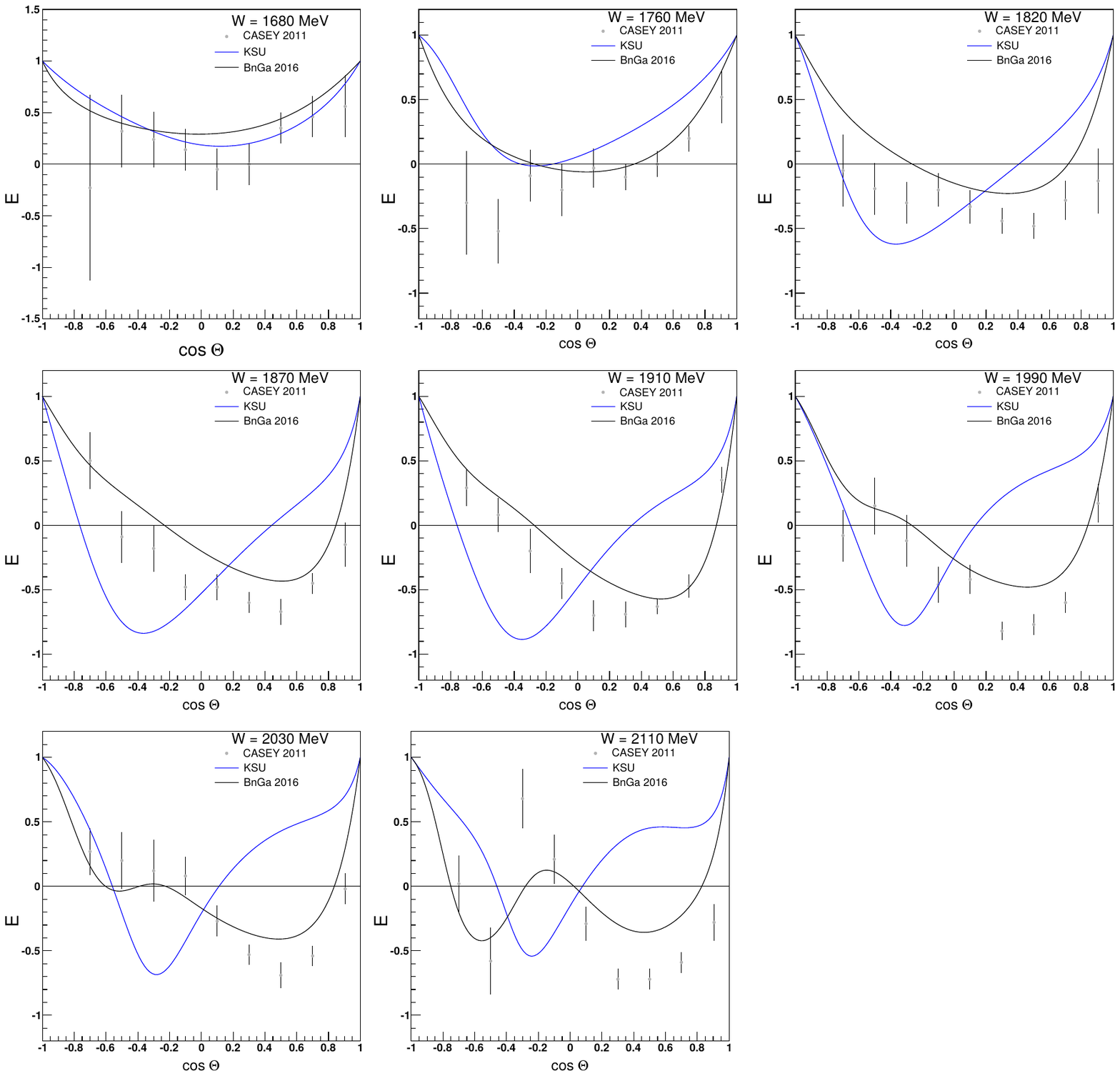}
	\caption{Fits to \printObs{} for \printReac{} at $W$ = 1680 to 2110~MeV.  See text for references.}
\end{figure}

\renewcommand{\ObsName}{F}
\renewcommand{\printObs}{$F$}
\noindent
\begin{figure}
	\centering
	\includegraphics[scale=\scaleFac,trim={\trimA} {\trimB} {\trimC} {\trimD},clip=true]{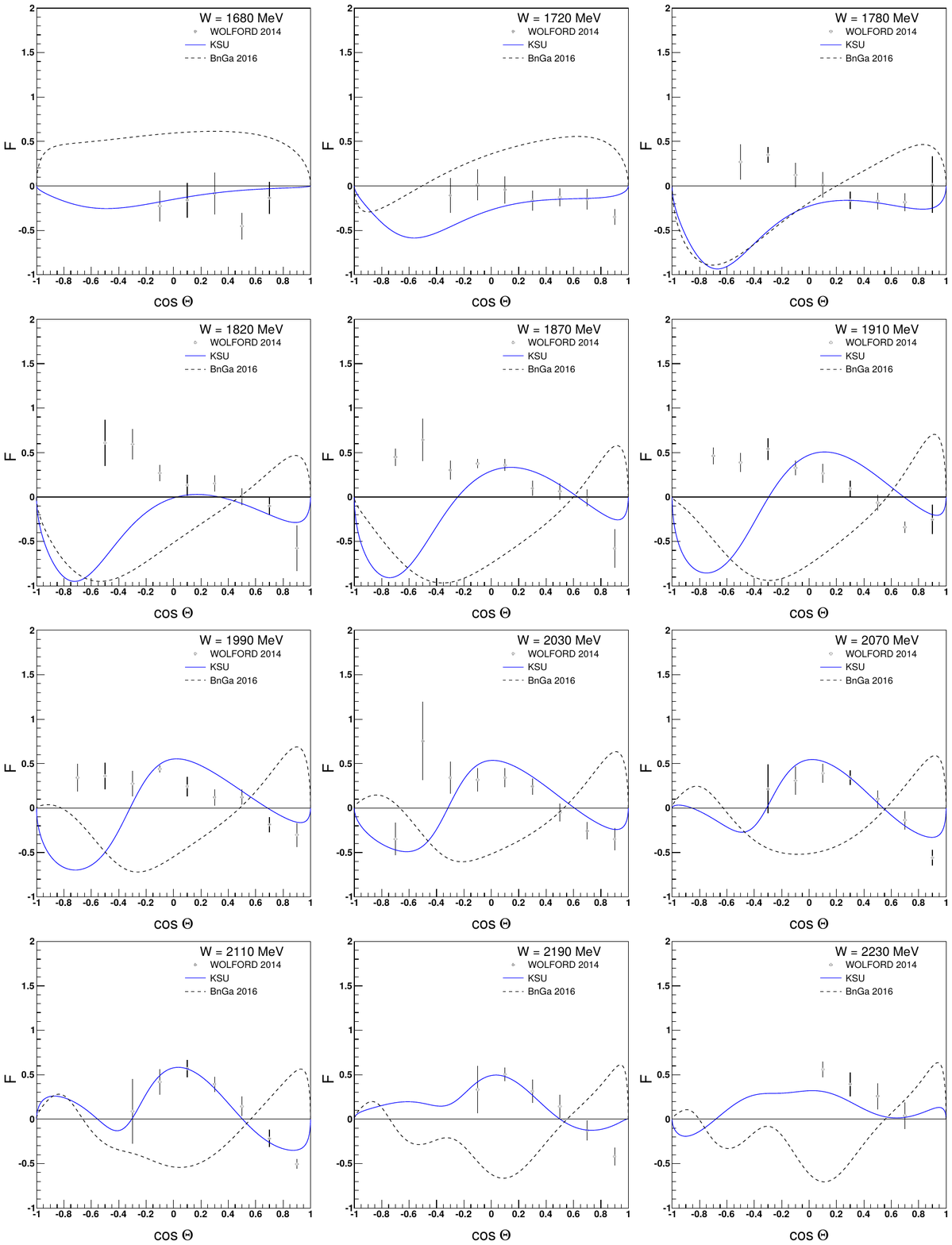}
	\caption{Fits to \printObs{} for  \printReac{} at $W$ = 1680 to 2230~MeV.  See text for references.}
\end{figure}

\renewcommand{\ObsName}{Cx}
\renewcommand{\printObs}{$C_x$}
\noindent
\begin{figure}
	\includegraphics[scale=\scaleFac,trim={\trimA} {\trimB} {\trimC} {\trimD},clip=true]{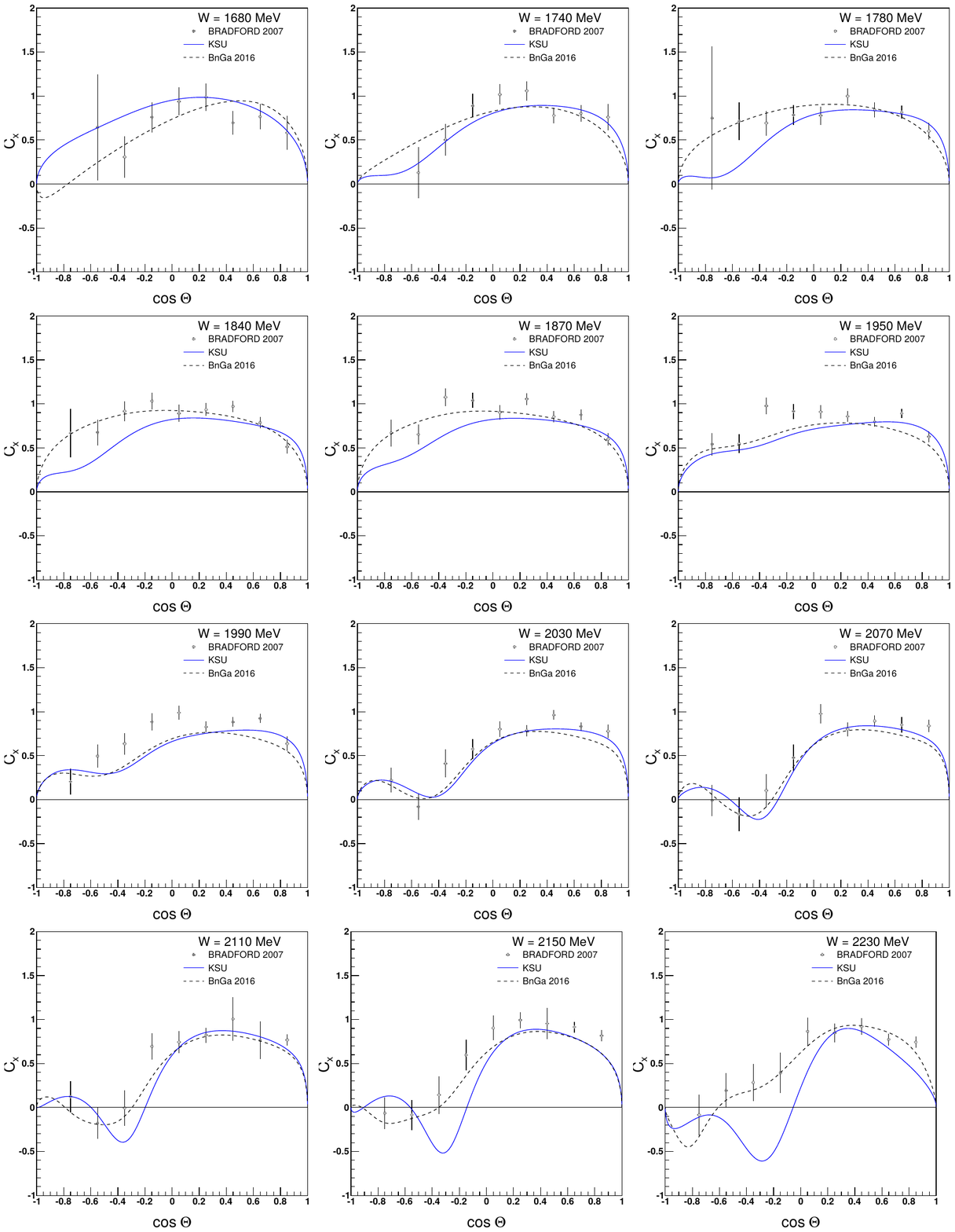}
	\caption{Fits to \printObs{} for \printReac{} at $W$ = 1680 to 2230~MeV.  See text for references.}
\end{figure}

\renewcommand{\ObsName}{Cz}
\renewcommand{\printObs}{$C_z$}
\noindent
\begin{figure}
	\includegraphics[scale=\scaleFac,trim={\trimA} {\trimB} {\trimC} {\trimD},clip=true]{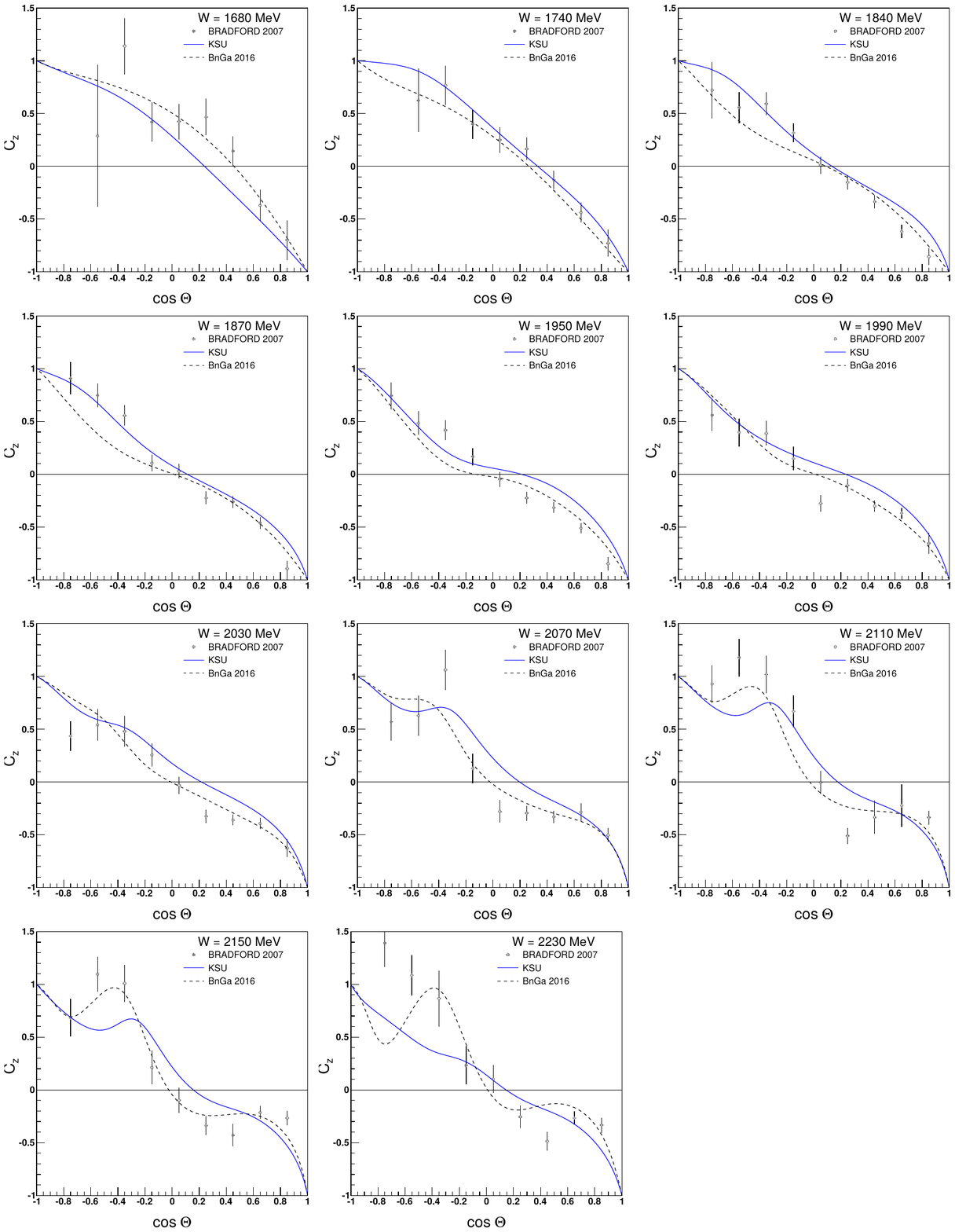}	
	\caption{Fits to \printObs{} for \printReac{} at $W$ = 1680 to 2230~MeV.  See text for references.}
\end{figure}

\renewcommand{\ObsName}{Ox}
\renewcommand{\printObs}{$O_x$}
\noindent
\begin{figure}
	\includegraphics[scale=\scaleFac,trim={\trimA} {88mm} {\trimC} {\trimD},clip=true]{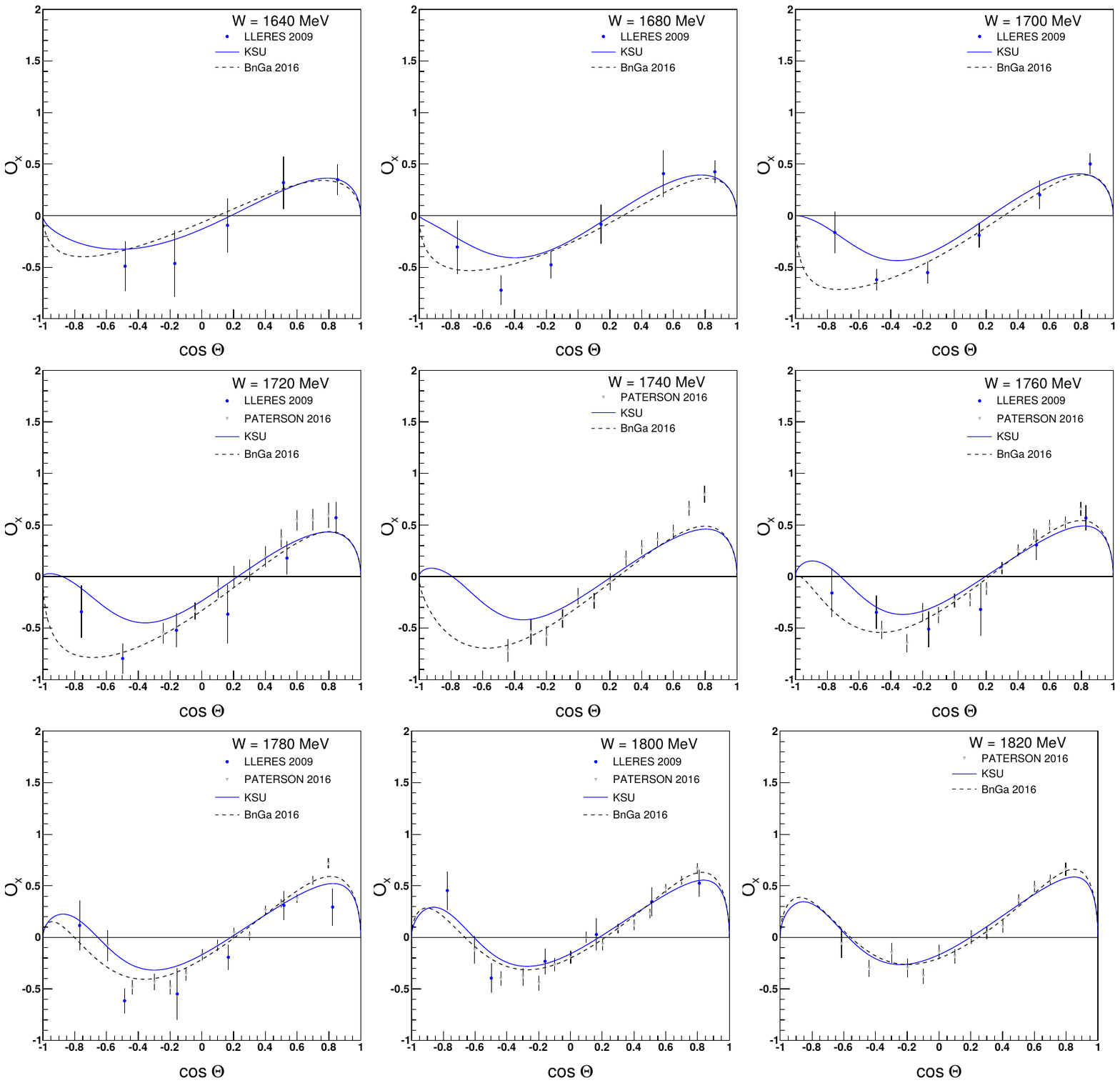}
	\caption{Fits to \printObs{} for \printReac{} at $W$ = 1640 to 1820~MeV.  See text for references.}
\end{figure}

\noindent
\begin{figure}
	\centering
	\includegraphics[scale=\scaleFac,trim={\trimA} {88mm} {\trimC} {\trimD},clip=true]{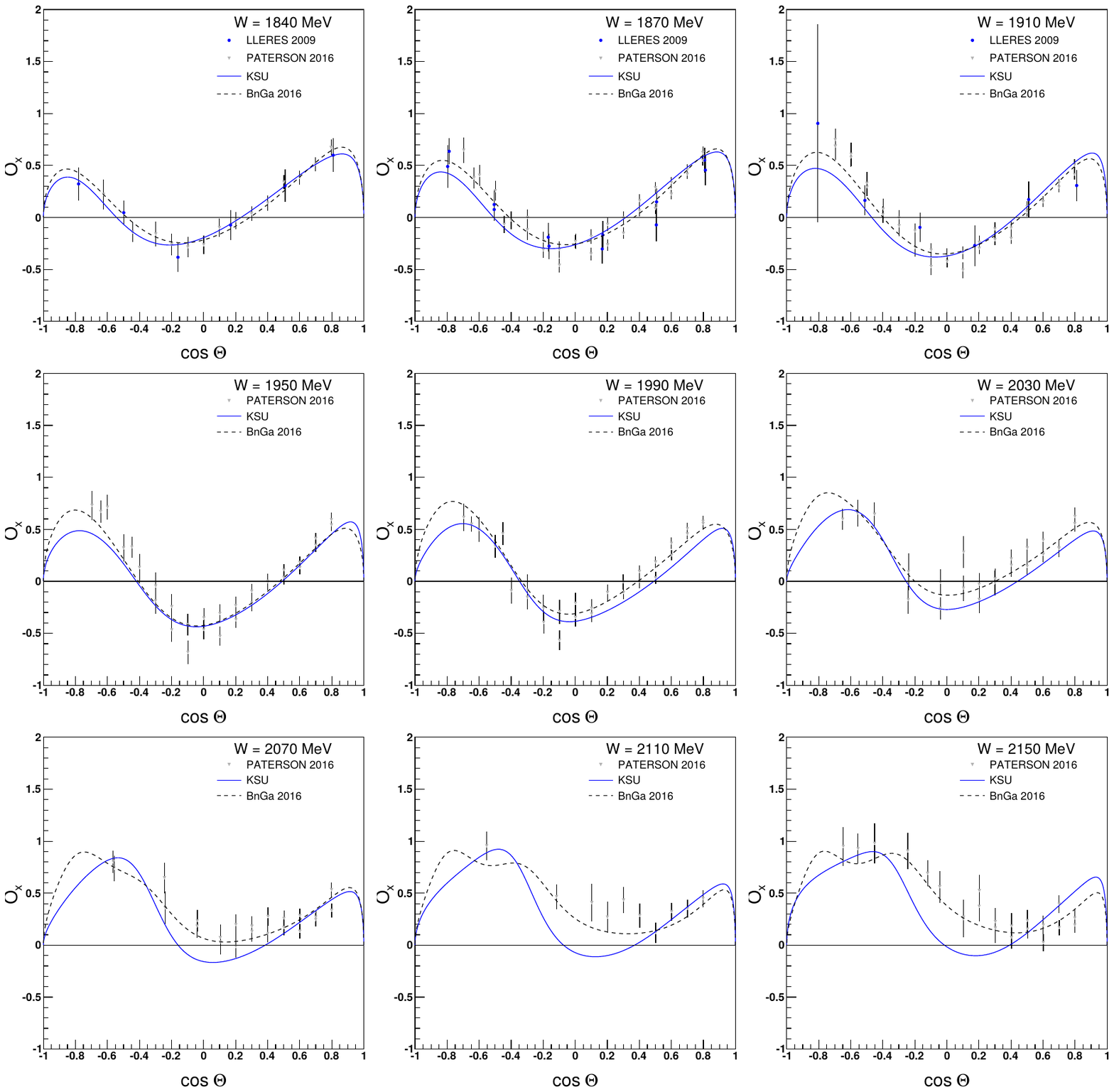}
	\caption{Fits to \printObs{} for \printReac{} at $W$ = 1840 to 2150~MeV.  See text for references.}
\end{figure}

\renewcommand{\ObsName}{Oz}
\renewcommand{\printObs}{$O_z$}
\noindent
\begin{figure}
	\includegraphics[scale=\scaleFac,trim={\trimA} {88mm} {\trimC} {\trimD},clip=true]{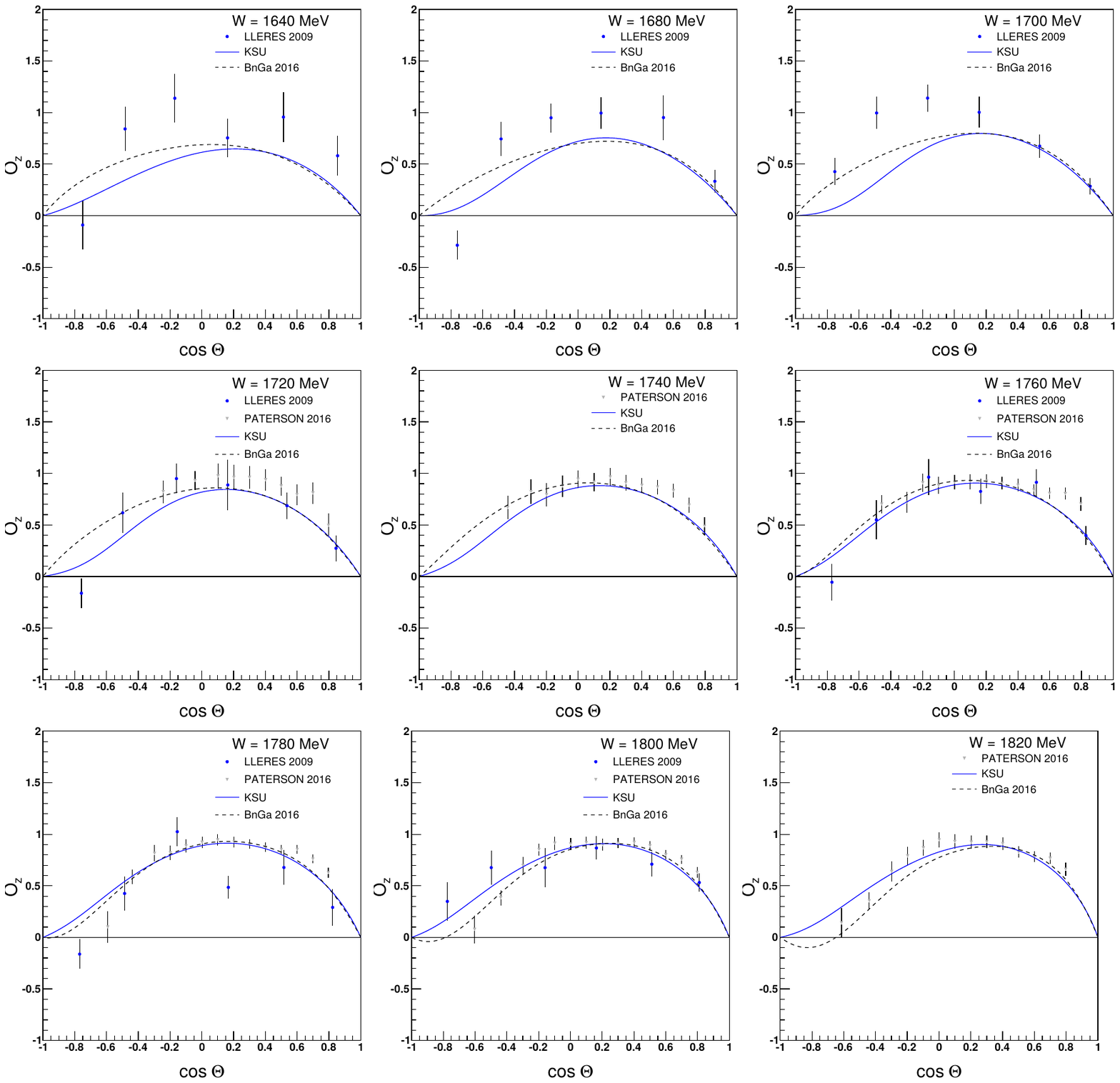}
	\caption{Fits to \printObs{} for \printReac{} at $W$ = 1640 to 1820~MeV.  See text for references.}
\end{figure}

\noindent
\begin{figure}
	\includegraphics[scale=\scaleFac,trim={\trimA} {88mm} {\trimC} {\trimD},clip=true]{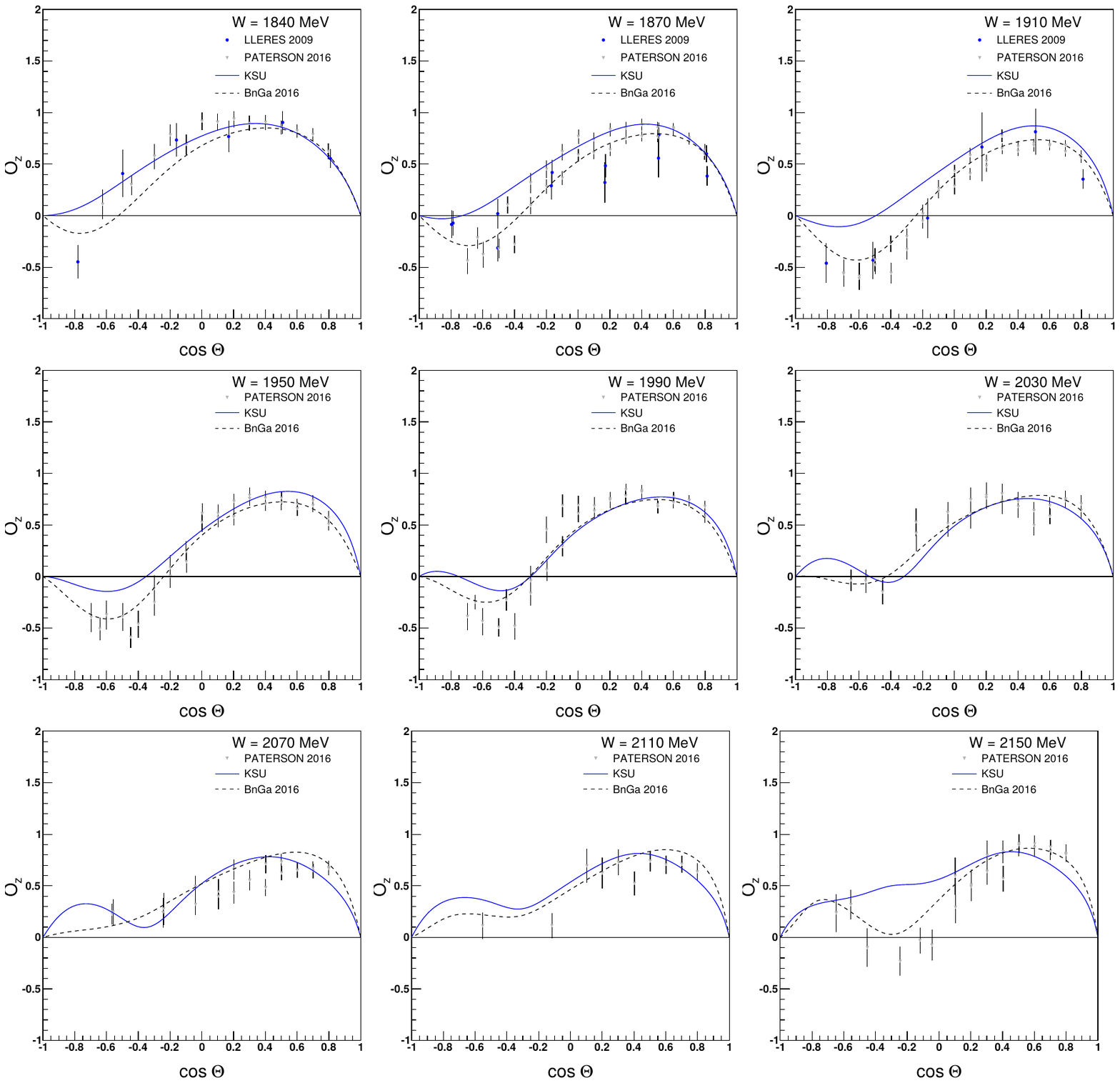}
	\caption{Fits to \printObs{} for \printReac{} at $W$ = 1840 to 2150~MeV.  See text for references.}
\end{figure}

\renewcommand{\ObsName}{Lx}
\renewcommand{\printObs}{$L_x$}
\noindent
\begin{figure}
	\includegraphics[scale=\scaleFac,trim={\trimA} {88mm} {\trimC} {\trimD},clip=true]{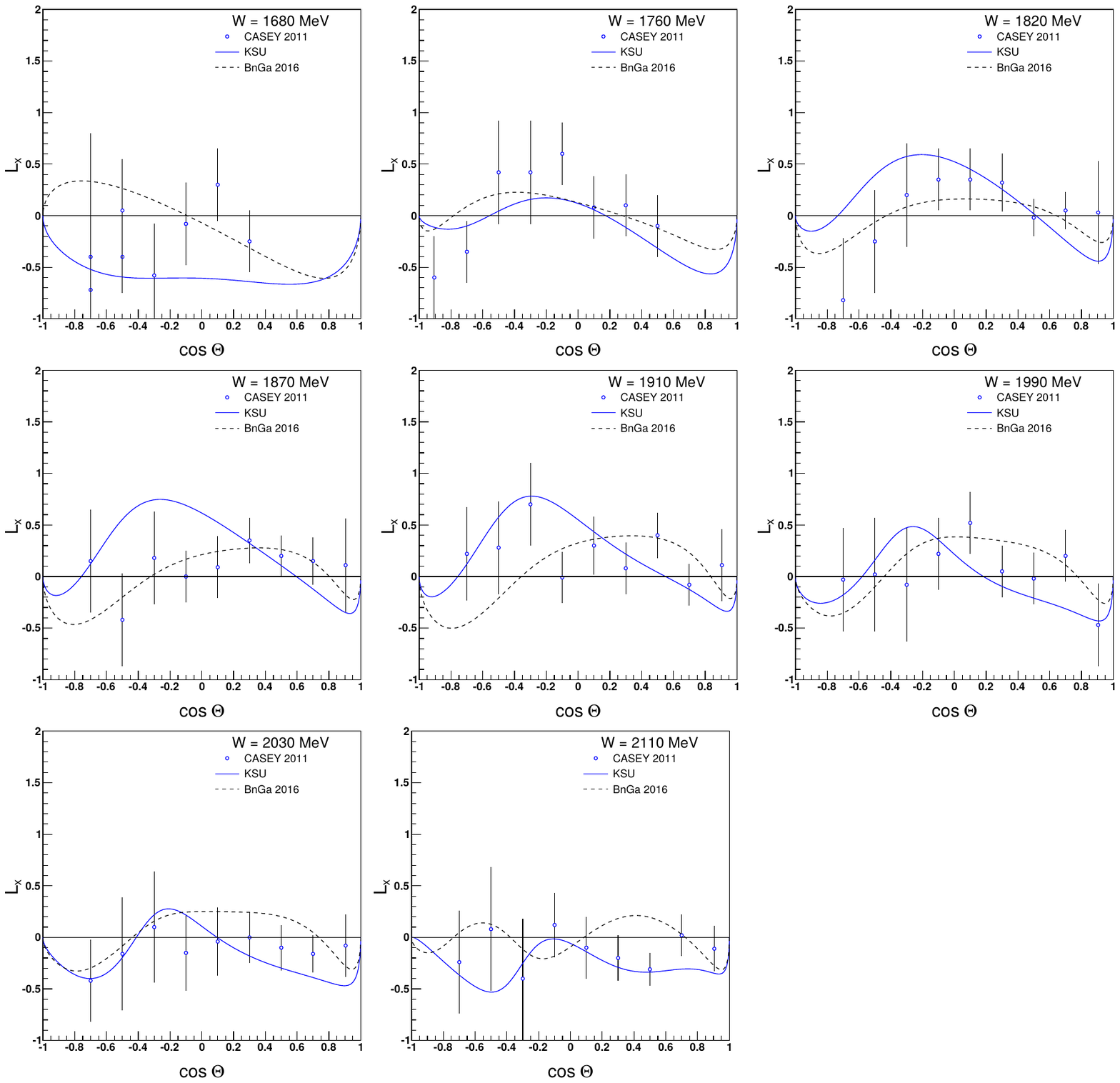}
	\caption{Fits to \printObs{} for \printReac{} at $W$ = 1680 to 2110~MeV.  See text for references.}
\end{figure}

\renewcommand{\ObsName}{Lz}
\renewcommand{\printObs}{$L_z$}
\noindent
\begin{figure}
	\includegraphics[scale=\scaleFac,trim={\trimA} {88mm} {\trimC} {\trimD},clip=true]{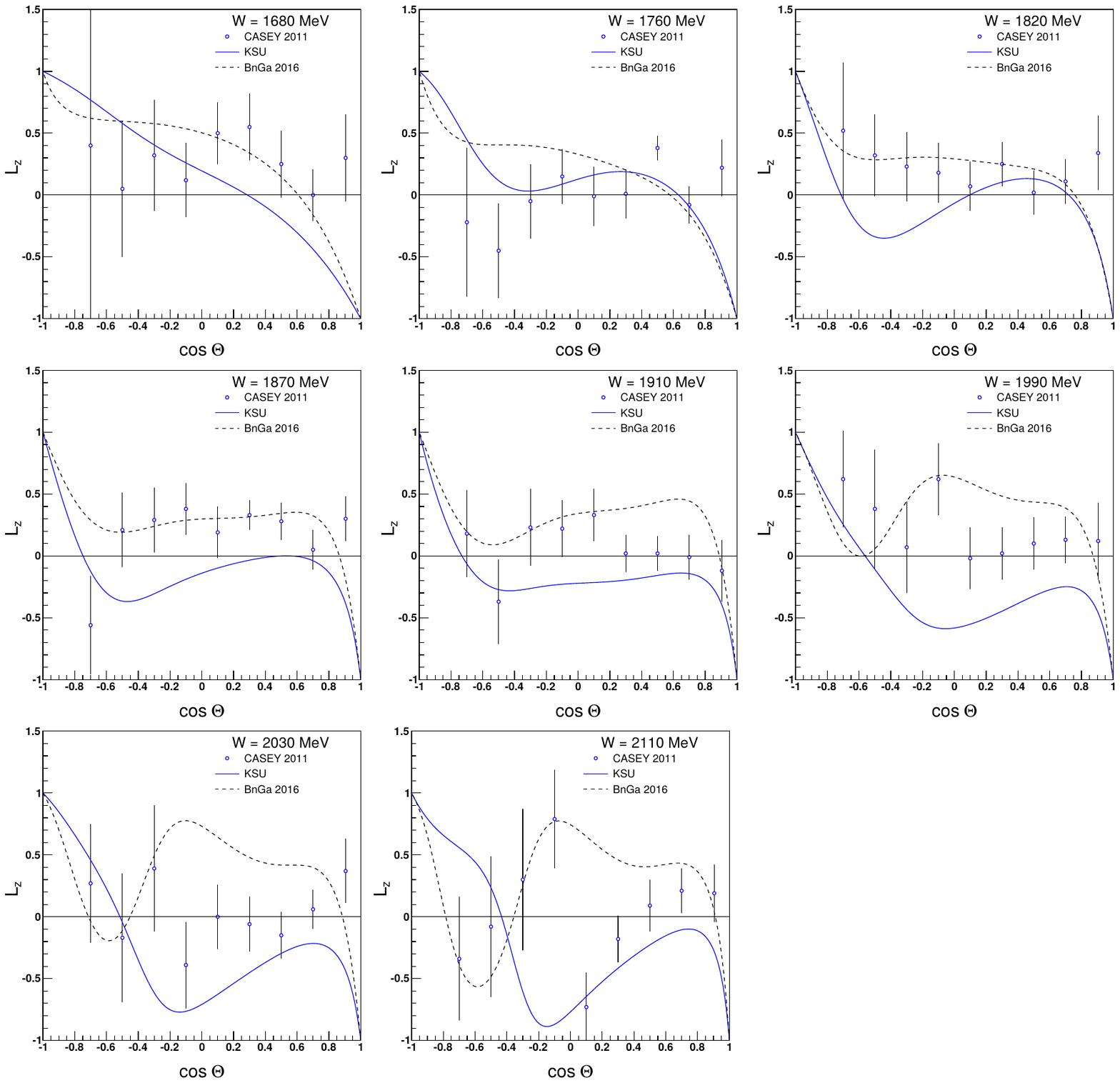}
	\caption{Fits to \printObs{} for  \printReac{} at $W$ = 1680 to 2110~MeV.   See text for references.}
\end{figure}

\renewcommand{\ObsName}{Tx}
\renewcommand{\printObs}{$T_x$}
\noindent
\begin{figure}
	\includegraphics[scale=\scaleFac,trim={\trimA} {\trimB} {\trimC} {\trimD},clip=true]{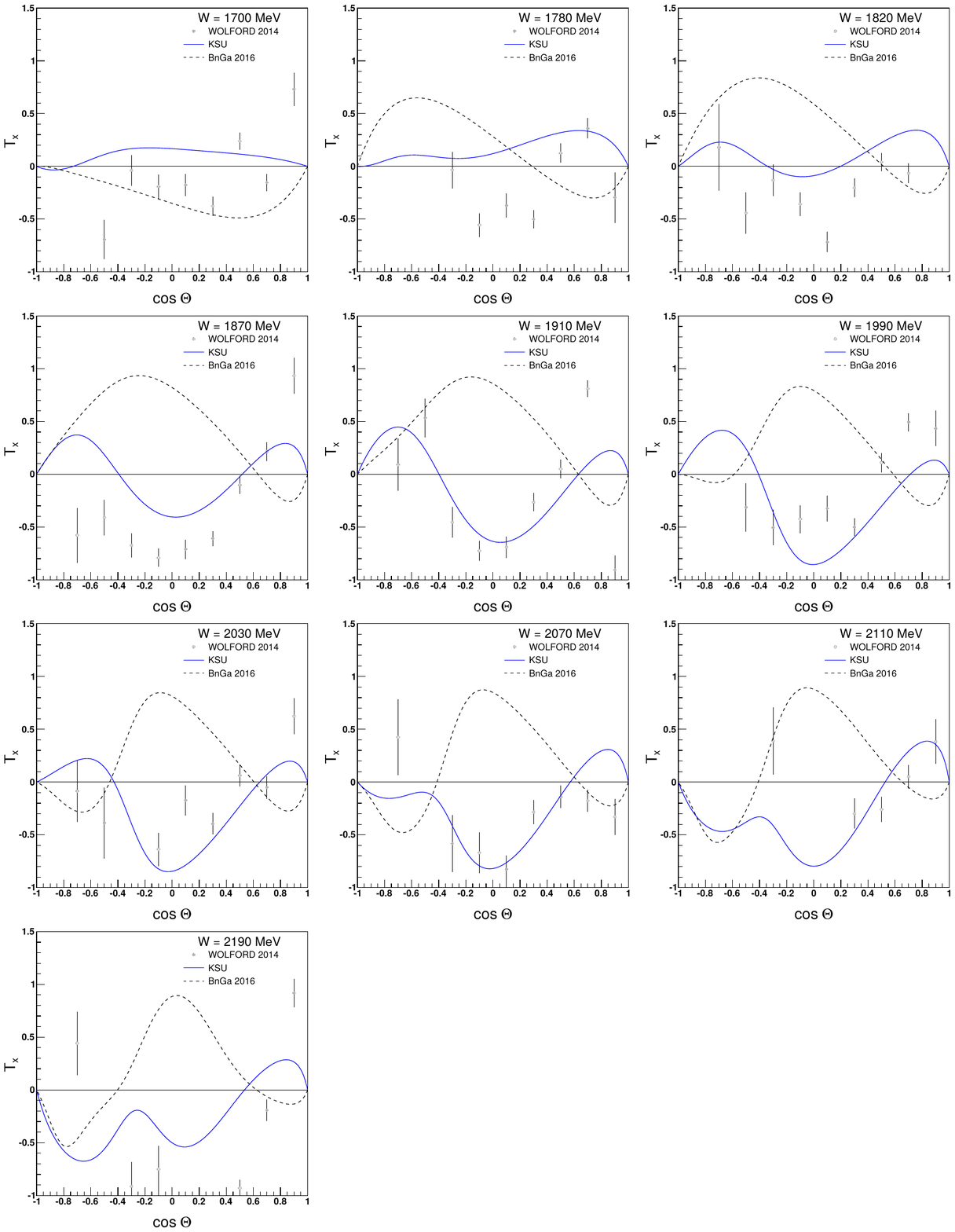}
	\caption{Fits to \printObs{} for  \printReac{} at $W$ = 1700 to 2190~MeV.  See text for references.}
\end{figure}

\renewcommand{\ObsName}{Tz}
\renewcommand{\printObs}{$T_z$}
\noindent
\begin{figure}	
	\includegraphics[scale=\scaleFac,trim={\trimA} {88mm} {\trimC} {\trimD},clip=true]{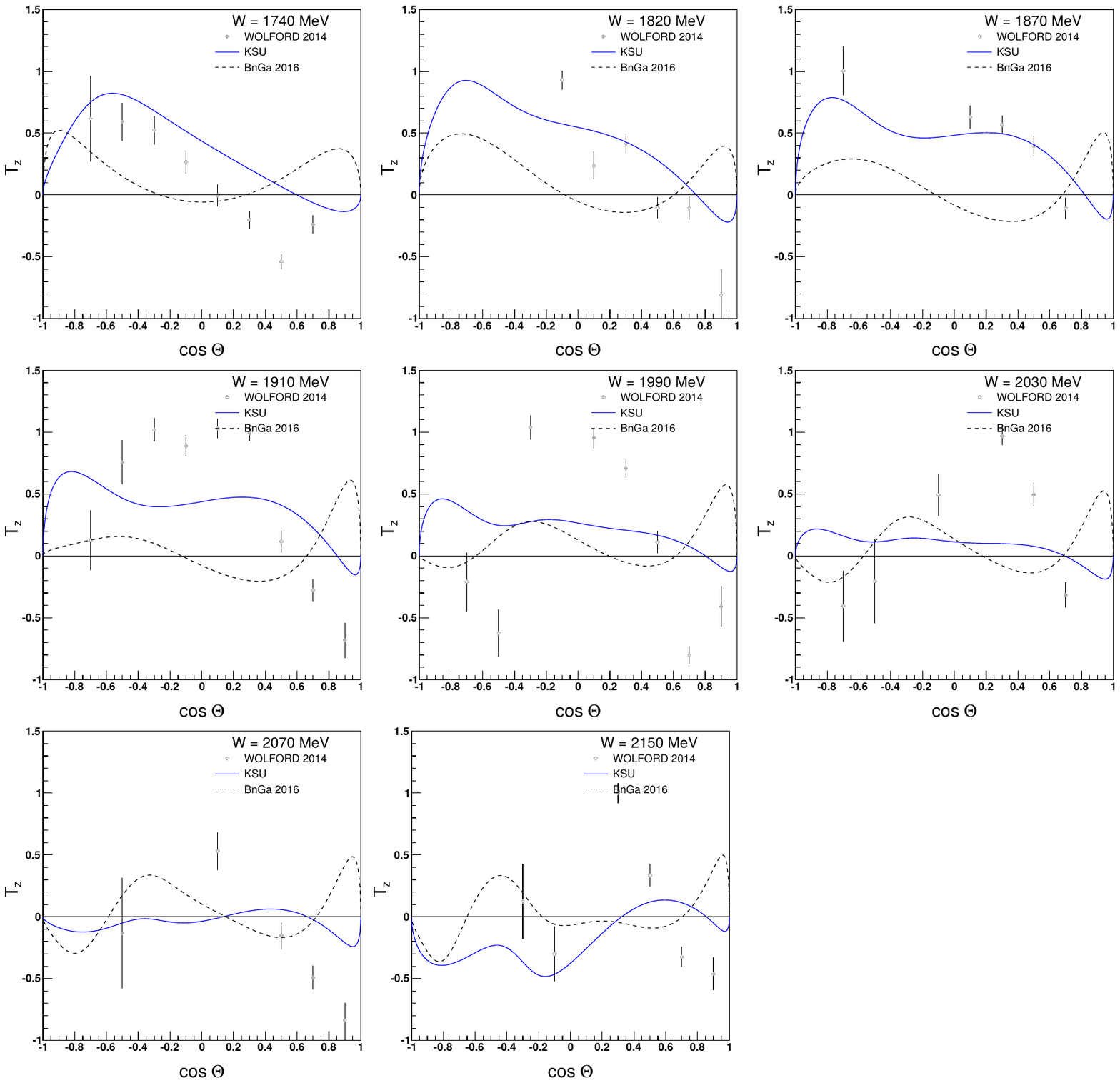}
	\caption{\label{GPKLLast} Fits to \printObs{} for \printReac{} at $W$ = 1740 to 2150~MeV.  See text for references.}
\end{figure}


\bibliography{basename of .bib file}

\end{document}